\documentclass[preprint,12pt,authoryear]{elsarticle}

\usepackage{graphicx}
\usepackage{amssymb}
\usepackage{amsmath}
\usepackage{lineno}
\usepackage{siunitx}
\usepackage{color}
\usepackage{natbib}
\usepackage{bm}
\usepackage{fullpage}
\usepackage[colorinlistoftodos,textsize=tiny]{todonotes}

\journal{Journal of Fluids and Structures}

\begin{document}

\begin{frontmatter}


\title{Assessment of low-altitude atmospheric turbulence models for aircraft aeroelasticity}

\author{Georgios Deskos}
\author{Alfonso del Carre}
\author{Rafael Palacios}
\ead{r.palacios@imperial.ac.uk}

\address{Department of Aeronautics, Imperial College London, United Kingdom}

\begin{abstract}
We investigate the dynamic response of flexible aircraft in low-altitude atmospheric turbulence. To this end, three turbulence models of increasing fidelity, namely, the one-dimensional von K{\'a}rm{\'a}n model, the two-dimensional Kaimal model and full three-dimensional wind fields extracted from large-eddy simulations (LES) are used to simulate ambient turbulence near the ground. Load calculations and flight trajectory predictions are conducted for a flexible high-aspect ratio aircraft, using a fully coupled nonlinear flight dynamics/aeroelastic model, when it operates in background atmospheric turbulence generated by the aforementioned models. Comparison of load envelopes and spectral content, on vehicles of varying flexibility, shows strong dependency between the selected turbulence model and aircraft aeroelastic response (e.g. 58\% difference in the predicted magnitude of the wing root bending moment between LES and von K{\'a}rm{\'a}n models). This is mainly due to the presence of large flow structures at low altitudes that have comparable dimensions to the vehicle, and which despite the relatively small wind speeds within the Earth boundary layer, result in overall high load events. Results show that one-dimensional models that do not capture those effects provide fairly non-conservative load estimates and are unsuitable for very flexible airframe design.

\end{abstract}
\begin{keyword}
Atmospheric turbulence \sep gust loads \sep geometrically-nonlinear structures \sep dynamic aeroelasticity \sep flight mechanics  \sep large-eddy simulation
\end{keyword}

\end{frontmatter}


\section{Introduction}\label{sec:Intro}
Atmospheric turbulence and its effects on aircraft flight has been a major concern of the aerospace sciences for more than a century \citep{HunsakerWilson1917}. The efforts to characterise the impact of atmospheric turbulence on aircraft loads can be traced back to the early development and consolidation of discrete gust and continuous turbulence models   \citep{Jones1940,Miles1956,zbrozek1965,Eichenbaum1971,Houbolt1973}. Today, airworthiness certification procedures (e.g. \cite{EASACS-232007}) have adapted guidelines and methodologies derived from both that early research and the accumulated experience of decades of in-service flight data. However, the very recent development of high-altitude pseudo satellites (HAPS) for an increasingly wide range of applications in remote sensing and communications \citep{TozerGrace2001,YangMohammed2010} is challenging the conventional understanding of the interactions between (nonlinear) aeroelasticity and atmospheric turbulence. The extreme operational demands and requirements of HAPS have resulted in more slender and much lighter platforms which inevitably renders them more flexible. This is true for example for stratospheric solar high-aspect ratio aircraft (NASA Helios, Boeing Oddyseus, Facebook Aquila, Airbus Zephyr, etc.) \citep{GonzaleEtAl2018}. Poor understanding of the low-altitude response of such very flexible aircraft (VFA) has previously led to poorly controlled flight dynamics and even vehicle loss \citep{NollEtAl2004}. Gaining physical insight into the aeroelastic interactions while quantifying the uncertainty of flight dynamics is considered key in further developing these technologies.

VFA aeroelasticity is characterised by large (geometrically-nonlinear) structural displacements and by strong interactions between the low-frequency structural modes and the vehicle flight dynamics. Conventional aircraft, on the other hand, present a certain degree of decoupling between the structure and the flight dynamics modes. Consequently, while aeroelastic analysis methods in the latter case are based on linearised frequency-domain solvers with potentially some higher-order corrections, the linear assumption brings limitations that can result in poor estimates in the response of VFA \citep{Cesnik2014}. \cite{Afonso2017} provide a comprehensive survey on the state of the art of VFA aeroelasticity with focus on commercial aviation. Indeed, while the current HAPS configurations are already relatively flexible and feature low-frequency structural modes, they are likely to be even more flexible as technology matures. 

\begin{figure*}[ht]
    \centering
    \includegraphics[width=\linewidth]{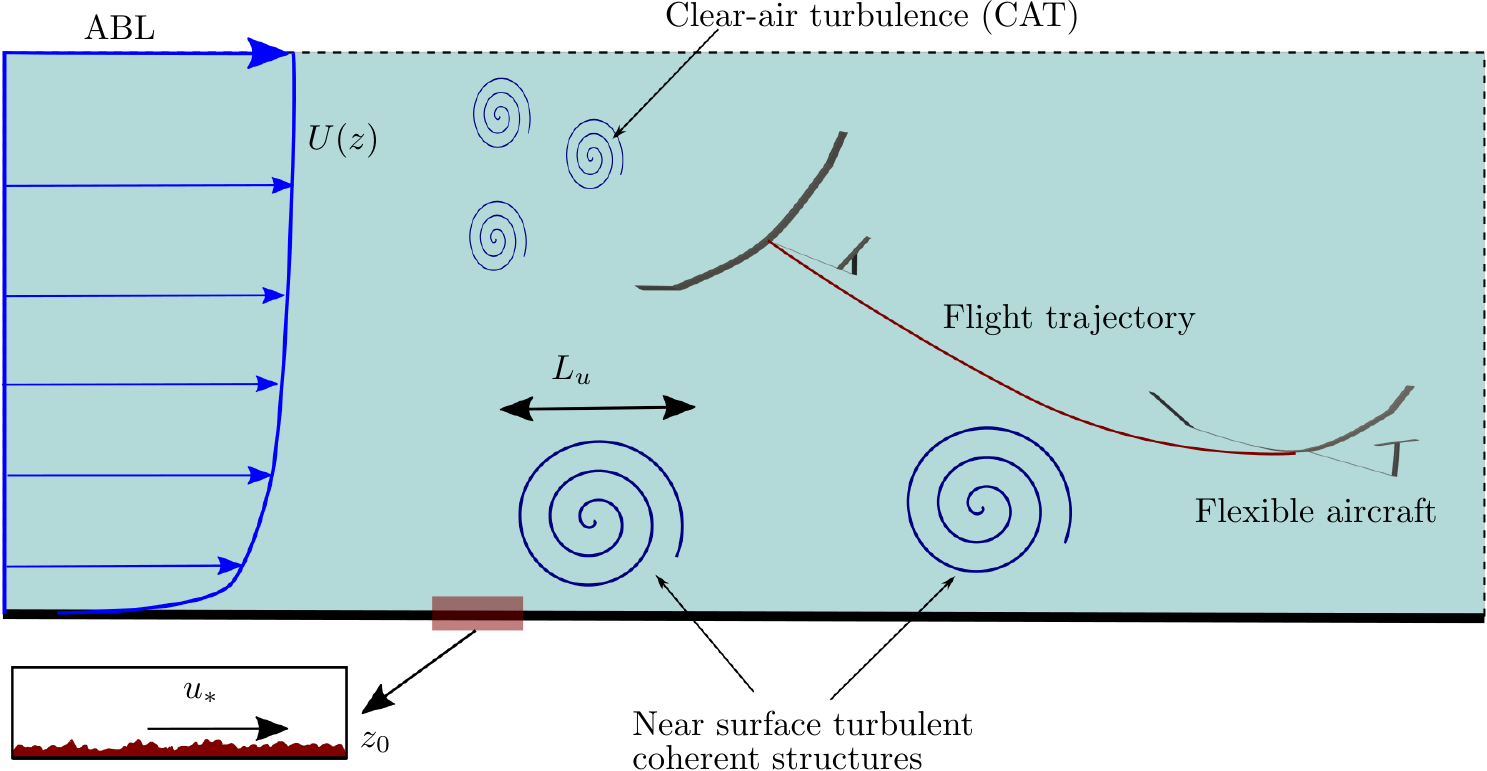}
    \caption{Schematic representation of the atmospheric turbulence experienced by the flexible aircraft.}
    \label{fig:SchematicHAPSoperatingEnv}
\end{figure*}

Continuous turbulence design criteria are often  based on the von K\'{a}rm\'{a}n spectrum with few empirical observations. Employing such approach to study atmospheric turbulence interacting with very flexible aircraft aeroelasticity poses two major issues. First, the von K{\'a}rm{\'a}n spectrum is derived from isotropic turbulence arguments, a condition which is rarely met in atmospheric turbulence, and never at low altitudes. In the lower part of the planetary boundary layer (surface layer), turbulence is strongly modulated by the presence of an aerodynamically rough surface and heat flux due to solar radiation. These effects combined together can significantly alter the flow dynamics and therefore deviate from the isotropic turbulence case implied by the von K\'{a}rm\'{a}n-like turbulence that is usually found in higher altitudes (e.g. CAT). The low-altitude operating environment of a generic VFA is shown schematically in figure \ref{fig:SchematicHAPSoperatingEnv}.
The near-surface turbulence coherent structures (e.g. high/low-speed streaks) will dominate the flight dynamics of such vehicle. The second shortcoming relates to the coupling between these large-scale turbulence structures (e.g. eddies) with the dynamics of very flexible aircraft. As high-wing deformations are expected under normal trim conditions, aircraft are prone to dynamic instabilities of significant magnitude \citep{Afonso2017,NollEtAl2004}. In particular, high-aspect ratio wings are expected to present a coupling between rigid-body and elastic motions since the frequency of the elastic modes are closer to those of the flight dynamics modes. This strong coupling has been well-documented in, e.g.,  \cite{Schmidt2012} and \cite{vonSchoorEtAl1990} for general aircraft aeroelasticity and, more specifically for high-aspect ratio aircraft, in \cite{Patil2001b}, \cite{TangDowell2004}, \cite{RaghavanPatil2009}, and \cite{StandfordBeran2013}, among others. To this day, a limited number of experimental studies on high-aspect ratio wings exists. They have shown however, that very flexible wings have a significantly different gust response characteristics under different deformations and load conditions \citep{LiuEtAl2016}. Inherently, additional analysis is required in order to gain insight into the complex interactions between wind gusts and aircraft aeroelasticity. This particular study focuses on investigating the effect of low-altitude turbulence on very flexible high-aspect ratio aircraft by first conducting a sensitivity analysis of atmospheric turbulence generation methods including the industry-standard von K\'{a}rm\'{a}n model as well as demonstrating the differences in the dynamic aeroelastic response between very flexible and stiff aircraft. 

To investigate the sensitivity of the different methods used to produce the continuous turbulent field on flexible aircraft gust loads, we examine the response of a representative very flexible aircraft model to three atmospheric turbulence input methods namely, the one-dimensional von K{\'a}rm{\'a}n, a two-dimensional stochastic formulation Kaimal model \citep{KaimalEtAl1972} enhanced with a spatial coherence model as well as full-three dimensional wind fields produced by means of large-eddy simulations \citep{Porte-AgelEtAl2001,Bou-ZeidEtAl2005,Deskos2019}. The paper starts with a brief introduction to atmospheric turbulence modelling techniques in Section \ref{sec:AtmosphericTurbulenceModelling}. The setups used to simulate the wind fields are presented in Section \ref{sec:CompTurbModels} together with comparing their statistics and respective velocity spectra. Subsequently, we make use of a fully-coupled computational aeroelasticity model to make time-domain predictions of the flight dynamics and loads. The results are presented in Section \ref{sec:AeroelasticityModel}. Results from the simulations including a sensitivity analysis for the three turbulence inputs and a cross-comparison between a flexible and a stiffer aircraft are presented in Section \ref{sec:AtmoAndAeroelast}. Finally, a summary together with some key findings are presented in Section \ref{sec:Conclusions}.

\section{Atmospheric turbulence modelling}\label{sec:AtmosphericTurbulenceModelling}
Within the atmospheric boundary layer (ABL), turbulence evolves in response to the local topography as well as the heating and cooling of the Earth's surface. The latter, categorises the ABL in three distinct states, namely, a neutral boundary layer (NBL), a convective boundary layer (CBL) and a stable boundary layer (SBL). In the neutral boundary layer, thermal effects remain unimportant and air parcels move adiabatically solely due to Coriolis forcing and surface friction. Thermal effects on the other hand are significant in the convective and stable boundary layers in which temperature stratification leads to a capping inversion (inversion layer). In the CBL, capping inversion acts as a lid and turbulence convection is dominated by large plumes (thermals) that transport and mix air away from the surface. The CBL state is often found following the sunrise. On the other hand, the SBL is characterised by a strong temperature stratification which typically leads to a shallower boundary layer thickness, strong wind shear and much smaller and weaker eddies than the neutral and convective cases \cite{Wyngaard2010}. Nonetheless, all three states (except for the convective case) are very rarely in a state of equilibrium and transition from one to another takes place multiple times during the day. However, provided the disparity between the ABL state evolution ($\sim$ hours) and the time spent by a VFA in the lower ABL ($\sim$ minutes), the assumption of ABL equilibrium can be considered a valid assumption.

In simulating atmospheric turbulence, two approaches can be generally considered. The first approach uses stochastic formulations to generate wind fields based on first and second-order statistics (spectral/co-spectra). Stochastic generation of wind fields has been an active area of research over the last decades \citep{Veers1988,Mann1994,Mann1998,MunozEsparzaEtAl2015}. For applications in wind engineering \citep{jeong2014,garcia2017}, the most common stochastic model is that of Kaimal \citep{KaimalEtAl1972,Veers1988}. The second technique on the other hand, utilises detailed numerical simulations to resolve the fluid flow equations (or a parametrisation of them) for a given temporal and spatial scale. Such a popular method is large-eddy simulation (LES). The idea underpinning LES lies in resolving the large energetic structures of the flow while modelling the smaller scales. LES produces three-dimensional wind field for a given computational domain and its quality depends on the resolution and modelling parameters. LES can be therefore considered as a high-fidelity model, as it resolves all relevant turbulent scales. Further details on both approaches are discussed next.

\subsection{Stochastic formulations for turbulence generation}\label{subsec:StochasticGeneration}
In analysing the dynamic response of an aircraft, three-dimensional isotropic turbulence models are often used to generate the turbulence input. They produce stochastic time signals for the three velocity components ($u,v,w$) by using either the Dryden or the von K{\'a}rm{\'a}n spectrum. The three-component signal is produced by passing band-limited white noise through forming filters obtained from spectral factorisation of the semi-empirical models \citep{LyChan1980,Campbell1986}. Uniform signals in transversal planes are subsequently generated in the area of interest creating a velocity field with perfect streamwise coherence, $\text{coh}_{uu}=\,$\num{1}. In this study we use the von K{\'a}rm{\'a}n model. For a nominal reference velocity $U_{ref}$, it is defined by the power spectral density,
\begin{equation}
   S_u(f) = \frac{2 \sigma_u^2 L_u}{\pi U_\textnormal{ref}} \cdot \frac{1}{\bigg[1+\bigg(1.339 L_u \frac{2 \pi f}{U_\textnormal{ref}} \bigg)^2 \bigg]^{5/6}}.
\end{equation}
for the longitudinal velocity and 
\begin{equation}
   S_{v,w}(f) = \frac{2 \sigma_{v,w}^2 L_{v,w}}{\pi U_\textnormal{ref}} \cdot \frac{1+\frac{8}{3}\bigg(2.678 L_{v,w} \frac{2 \pi f}{U_\textnormal{ref}} \bigg)^2}{\bigg[1+\bigg(2.678 L_{v,w} \frac{2 \pi f}{U_\textnormal{ref}} \bigg)^2 \bigg]^{11/6}}.
\end{equation}
for the lateral velocities. As our interest is in the low-altitudes, independent turbulence length scales $L_u,L_v,L_w$ are assumed in three directions. They are chosen as in \cite{DoDMILHDBK1997} so that $2L_w=h$ and $L_u=2L_v=\,h/(0.177+0.000823 h)^{1.2}$, where $h$ is the altitude (in feet). The corresponding turbulence intensities are $\sigma_u,\sigma_v,\sigma_w$.
For the simulations, a time-signal of \SI{250}{\second} is generated and subsequently convected into a full wind field at speed $U_{ref}$ by means of the ``frozen turbulence'' hypothesis. 

A more accurate representation of low-altitude atmospheric turbulence can also be sought via the Kaimal model. The Kaimal model can be used to make predictions of the wind field during the neutral boundary layer state using empirically calibrated constants. In the present study, Kaimal is used in conjunction with a spatial coherence model for the streamwise component of the velocity field, $u(y, z, t)$, known as the International Electrotechnical Commission (IEC) exponential coherence model \citep{IEC-61400-1},
\begin{equation}
    \text{coh}_{uu}(f,r,U_\textnormal{ref})=\exp \bigg[ -a \sqrt{\bigg(\frac{r f}{U_\textnormal{ref}}\bigg)^2+\bigg(\frac{r b}{L_u}\bigg)^2}\bigg]
\end{equation}
where $a=\,$\num{10} and $b=\,$\num{0.12} are the coherence decrement and offset parameters, $L_u$ the streamwise turbulence lengthscale, $U_\textnormal{ref}$ is the mean wind speed at the reference height, $f$ denotes frequency and $r$ is a radial distance in a plane vertical to the streamwise direction. To generate the 2D wind field we make use of the open-source code TurbSim \citep{TurbSim}. As TurbSim outputs planar fields of the velocity field as a function of time we use again Taylor's frozen turbulence hypothesis, $u(x,t)=u(0, x-t U)$, to generate a full three-dimensional velocity field.

\subsection{High-fidelity turbulent flows via large-eddy simulation}\label{subsec:LES}
Large-eddy simulations (LES) of atmospheric turbulence are conducted using our in-house solver \texttt{WInc3D} \citep{DeskosEtAl2018b,Deskos2019}. It is an explicit LES framework which solves the unsteady, incompressible, filtered Navier-Stokes equations using their skew-symmetric form 
\begin{equation}
\frac{\partial \widetilde{u}_i}{\partial t}+\frac{1}{2}\bigg(\widetilde{u}_j\frac{\partial \widetilde{u}_i}{\partial x_j} + \frac{\partial \widetilde{u}_i \widetilde{u}_j}{\partial x_j}\bigg)=-\frac{1}{\rho}\frac{\partial \widetilde{p^*}}{\partial x_i}-\frac{1}{\rho}\frac{\partial p_\infty}{\partial x_i}
-\frac{\partial \tau_{ij}}{\partial x_j}, \quad i,j=\{1,2,3\},
\label{eq:NS}
\end{equation}
\begin{equation}
\frac{\partial \widetilde{u}_i}{\partial x_i}=0,
\label{eq:Continuity}
\end{equation}
where $\widetilde{p^*}=\widetilde{p}+1/3 \widetilde{u_i} \widetilde{u_i}$, and $\widetilde{u}_i$ are the filtered components of the modified pressure and velocity fields, respectively, and $\rho$ is the fluid density. The indices $(1,2,3)$ correspond to a $(x,y,z)$ coordinate system. The sub-filter stresses $-\partial_j \tau_{ij}$ (residual terms from the filtering process) are calculated using the standard Smagorinsky model \citep{Smagorinsky1963} 
\begin{equation}
\tau_{ij}=-2(C_S \Delta)^2 |\widetilde{S}| \widetilde{S}_{ij}, \quad \widetilde{S}_{ij}=\frac{1}{2}\bigg(\frac{\partial \widetilde{u}_i}{\partial x_j} + \frac{\partial \widetilde{u}_j}{\partial x_i} \bigg),
\label{eq:SmagorinskyModel}
\end{equation}
where $C_S$ is the Smagorinsky constant which is corrected near the wall using the \cite{MasonThomson1992}
\begin{equation}
C_S=\bigg(C_0^n + \bigg \{ \kappa \bigg(\frac{z}{\Delta}+\frac{z_0}{\Delta} \bigg)\bigg \}^{-n} \bigg)^{-1/n}.
\end{equation}
where $z_0$ is the roughness lengthscale, $\Delta=\sqrt[3]{\Delta x \Delta y \Delta z}$ the grid size and choosing $(C_0,n)=\,(0.14,3)$. Finally, $-\partial_i p_\infty $ is a constant pressure gradient applied to flow to reproduce the effect of geostrophic forcing. It is worth noting that in equation \eqref{eq:NS} the viscous term of the Navier-Stokes equation has been neglected. This is due to the high-Reynolds number and the fact that the mesh is coarse enough so that the near-ground viscous layer will not have any impact on the calculations. Another implication of being under-resolved in the near-ground surface is that applying a no-slip boundary condition is devoid of any meaning. Instead, we simulate the bulk effect of velocity shear by replacing it with slip velocity and imposing a local shear stress through a wall-stress model. Commonly, the wall-stress model of \cite{Moeng1984} is used by considering the Monin-Obukhov similarity (MOS) theory and in this study we employ the recent formulations of \cite{Bou-ZeidEtAl2005},
\begin{equation}
\tau^{\text{wall}}(x,y)=\tau_w(x,y)\frac{\widehat{\widetilde{u}}_i(x,y,\Delta z/2)}{\sqrt{\widehat{\widetilde{u}}_x^{\,2}(x,y,\Delta z/2)+\widehat{\widetilde{u}}_y^{\,2}(x,y,\Delta z/2)}},
\end{equation}
and
\begin{equation}
\tau_w(x,y)=-\left[\frac{\kappa}{\ln \left(\frac{\Delta z /2}{z_0}\right)} \right]^2 \left[\widehat{\widetilde{u}}_x^{\,2}(x,y,\Delta z/2)+\widehat{\widetilde{u}}_y^{\,2}(x,y,\Delta z/2)\right].
\label{eq:WallStress}
\end{equation}
in which the horizontal-averaged boundary condition is replaced by a twice-filtered velocity. In addition, we make use of periodic boundary conditions in lateral directions and free/partial-slip ($\partial_3 \widetilde{u}_i=\widetilde{u}_3=0$ for $i \in {1,2}$) for the bottom $z=0$, and the top boundary $z=H$, respectively. To solve the governing equations, sixth-order compact finite-difference schemes are used \citep{LaizetLamballais2009} on a Cartesian mesh in a half-staggered arrangement (the same mesh is used for the three velocity components $(u_1,u_2,u_3)=\,(u,v,w)$, with a different mesh used for pressure $p$), pressure correction is achieved by a direct spectral solver, while for the time integration a low-storage third-order Runge-Kutta scheme is employed. Additionally, explicit filtering is also applied to all fields and at each time step using discrete sixth-order filters \citep{GaitondVisbal1998}. Finally, massive parallelisation of the numerical solver is achieved using MPI and an efficient 2D pencil domain decomposition approach \citep{LaizetLi2011}. The current implementation has been run on over $\mathcal{O}(10^5)$ number of processors and in multiple platforms and has consistently shown excellent scaling properties.

\section{Comparing turbulence generation methods}\label{sec:CompTurbModels}
\begin{figure*}[ht]
    \centering
    \includegraphics[width=\linewidth]{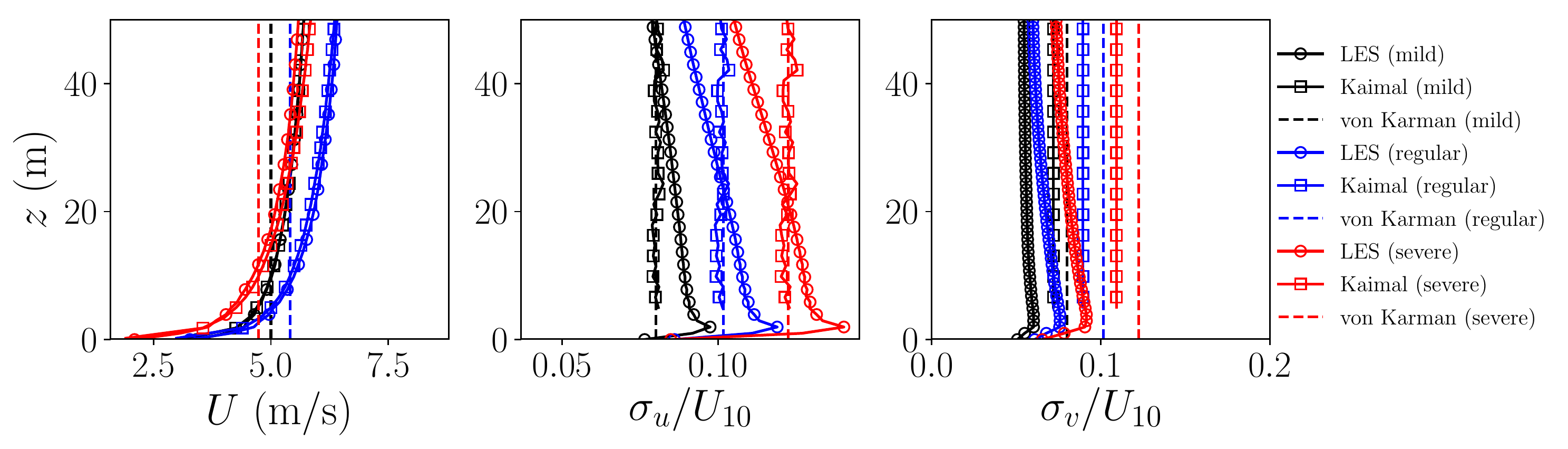}
    \caption{Mean velocity and turbulence intensity profiles for the three cases}
    \label{fig:ABLStatistics}
\end{figure*}

\begin{figure*}[ht]
    \centering
    \includegraphics[width=\linewidth]{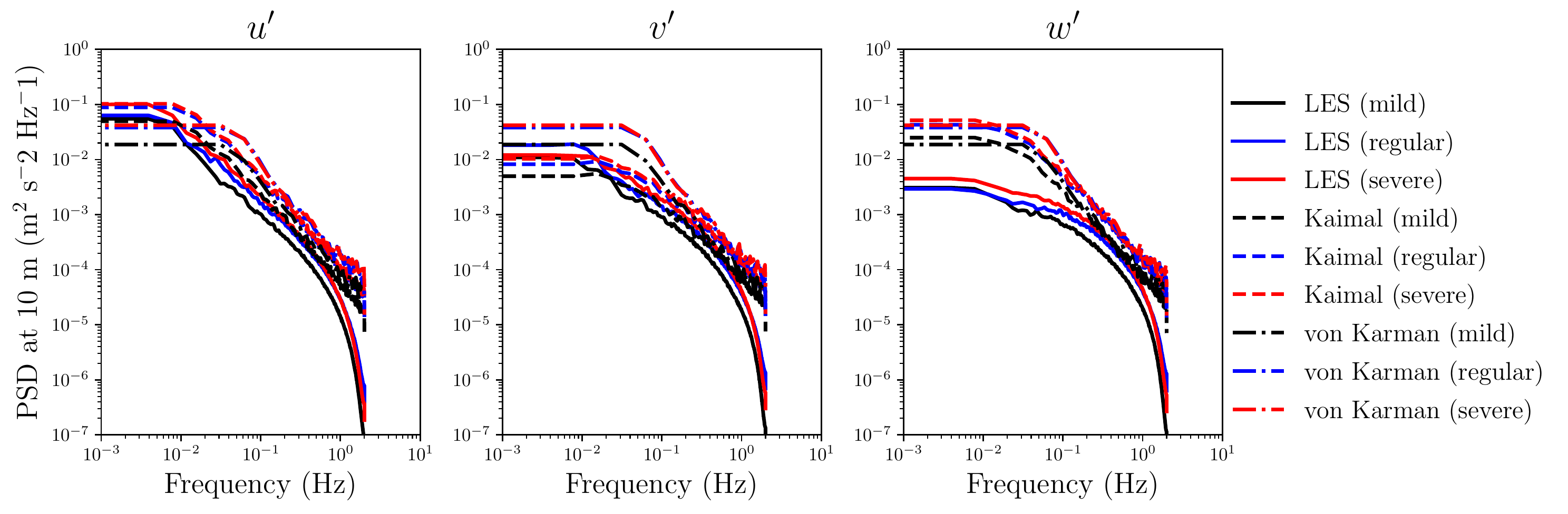}
    \caption{Power spectral density of velocity fluctuations for each velocity component ($u^\prime$, $v^\prime$ and $w^\prime$) and all three models (1D von K{\'a}rm{\'a}n, 2D Kaimal and 3D LES).  }
    \label{fig:ABLSpectra}
\end{figure*}

For a robust comparison between the different synthetic turbulence models (von K{\'a}rm{\'a}n, Kaimal, and LES), we must first ensure that they all yield statistically similar results. This includes both the time-averaged velocity profiles as well as velocity spectra. Here, the neutral atmospheric boundary layer on a flat Earth is considered with three levels of surface roughness, $z_0=\,$\SIlist{0.0001;0.001;0.01}{\meter}. This defines three flow conditions that will be referred to as \emph{Mild}, \emph{Regular} and \emph{Severe}, respectively, and correspond to surface friction velocities $u_*=\,$\SIlist{0.174;0.235;0.274}{\meter \per \second}. The friction velocities have been selected such that the mean velocity at altitude \SI{10}{\metre} attains a value of approximately $U_{10} \approx \,$\SI{5}{\metre \per \second} in all cases. Further details are provided in the table \ref{tab:Cases1} below. 
\begin{table}[!ht]
\caption{\label{tab:Cases1} Atmospheric turbulence flow parameters}
\centering
\begin{tabular}{lcccc}
\hline
Name& $z_0$ [m]& $u_*$ [m/s] &$U_{10}$ [m/s]& $\mathcal{I}$ \%\\\hline
{\it Mild}     & 0.0001 &  0.174  & 5.05 & 8.87   \\
{\it Regular}  & 0.001  &  0.235  & 5.41 & 10.74  \\
{\it Severe}   & 0.01   &  0.274  & 4.74 & 12.73  \\
\hline
\end{tabular}
\end{table}
Before proceeding to presenting the results it is also instructive to report the simulation configurations and some key parameters used for generating the data. Starting with large-eddy simulations, we consider a region of $L_x \: \times \: L_y \: \times  \:L_z=\SI{1000}{\metre} \: \times \: \SI{500}{\metre} \: \times \: \SI{500}{\metre}$ discretised by $N_x \times N_y \times N_z=\,1024 \times 512 \times 513$ mesh nodes. The time step was selected to be equal to $\Delta t=\,$\SI{0.05}{\second} so that a Courant-Friedrichs-Lewy (CFL) number of approximately $\num{0.25}$ is maintained throughout.
Each simulation was run for 50 large-eddy turnover times ($\delta/u_*$,  where $\delta=\,$\SI{1000}{\metre} is the boundary layer height) a spin-up time that has been previously used to obtain converged statistics \citep{Bou-ZeidEtAl2005,XieEtAl2015}. Restricted by the CFL number, the  total number of time steps required during the precursor simulation amounts to \num{2} million. All three cases were run using \num{4096} processors and required approximately \num{1.5} million CPU hours each. For the simulations involving the aeroelastic coupling, a small number of snapshots were extracted (every 3 seconds) and interpolated to obtain wind fields for a total of 10 minutes. For the Kaimal model a smaller mesh of $N_y \: \times \: N_z=\,$\num{32} $\times$ \num{32} nodes was used for a domain of $L_y \: \times  \:L_z=\SI{50}{\metre} \: \times \: \SI{50}{\metre}$, obtaining a resolution of approximately \SI{1.5}{\metre}. The calculations were made for a total analysis time of \SI{250}{\second} using a time step $\Delta t=\,$\SI{0.25}{\second}. To reproduce the desired results TurbSim \citep{TurbSim} was used in a customised mode by applying a logarithmic profile and setting the turbulence level at the characteristic altitude, surface frictions and coherence parameters so that we closely match the LES results. Lastly, for the von K{\'a}rm{\'a}n model, the standard deviations of the three velocity components $\sigma_u$, $\sigma_v$ and $\sigma_w$ together with the length scale $L_u$ extracted from the LES simulations were used to generate zero-mean time series before added to the background characteristic velocity $U_{10}$. A comparison of the vertical profiles and power spectral densities (PSD) are shown in figure \ref{fig:ABLStatistics} and \ref{fig:ABLSpectra}, respectively. The vertical mean velocity profiles agree well for the Kaimal and LES data whereas the von K\'{a}rm\'{a}n can only match the velocity at the reference altitude.
Larger discrepancies can be found between the the models in the streamwise and transverse velocity variances. In particular, LES predicts a linear decrease of the streamwise variance, a result that agrees with the observed data as well as theoretical scaling laws \citep{Wyngaard2010}, whereas the Kaimal and von K\'{a}rm\'{a}n models yield uniform distributions for all components.
Similarly, PSDs from the three models, shown in figure \ref{fig:ABLSpectra}, appear to collapse together for the streamwise component but exhibit larger discrepancies in the lateral ones. Specifically, in the vertical velocity component PSDs, LES data plots significantly deviate from those of Kaimal and von K\'{a}rm\'{a}n. Those two models assume flow isotropy which results in nearly identical PSDs over the $v'$ and $w'$ components.
\begin{figure*}[!ht]
    \centering
    \includegraphics[width=\linewidth]{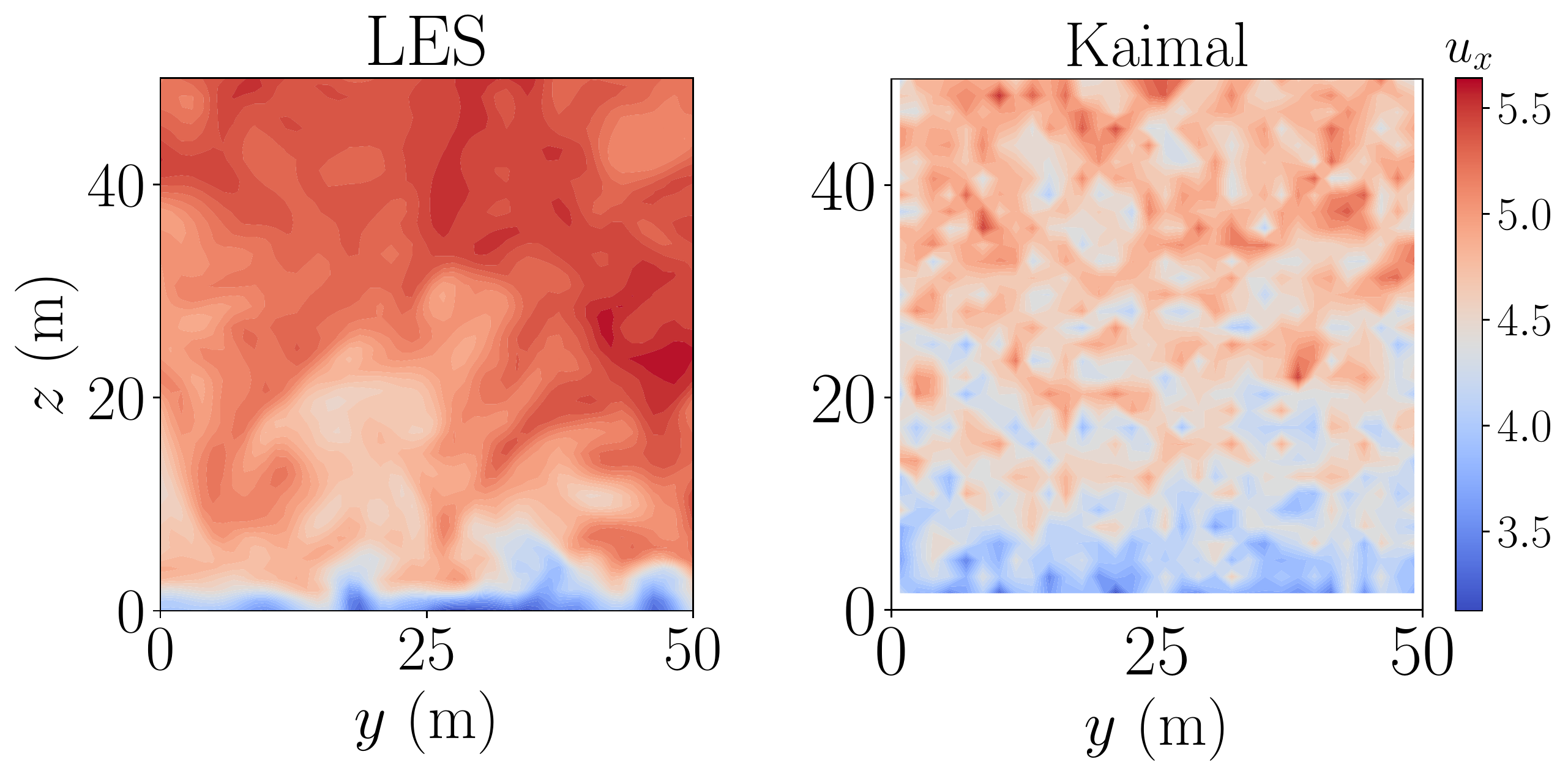}
    \caption{Instantaneous snapshots of a \emph{mild} streamwise velocity, $u(t)$, for the Kaimal and LES models.}
    \label{fig:InstaProfiles}
\end{figure*}
Nevertheless, while our comparison of the flow characteristics is ended here by comparing at most second-order statistics (velocity variance, spectra) more differences between the LES, which produces a non-gaussian distribution of velocity, and the stochastically-generated turbulence can be further sought. This is for example demonstrated qualitatively in figure \ref{fig:InstaProfiles}. The two snapshots of the LES and Kaimal, respectively, are both taken from the \emph{Mild} case and exhibit large differences in the formation of the near-wall large coherent structures. LES flow structures consist of large contra-rotating streamtubes generating large structures which extend as high as \SI{25}{\metre} while Kaimal consists of smaller structures extending only a few metres along the y-direction. The von K{\'a}rm{\'a}n model on the other hand is not plotted as only a uniform distribution will be assumed in a 2D planar snapshot. These qualitatively observed differences between the three models which are not captured by either the flow statistics will be re-assessed later on during the aeroelastic simulations in Section \ref{sec:AtmoAndAeroelast}.

\section{Nonlinear time-domain aeroelastic solver}\label{sec:AeroelasticityModel}
\begin{figure*}[bt]
\centering
\includegraphics[width=.8\textwidth]{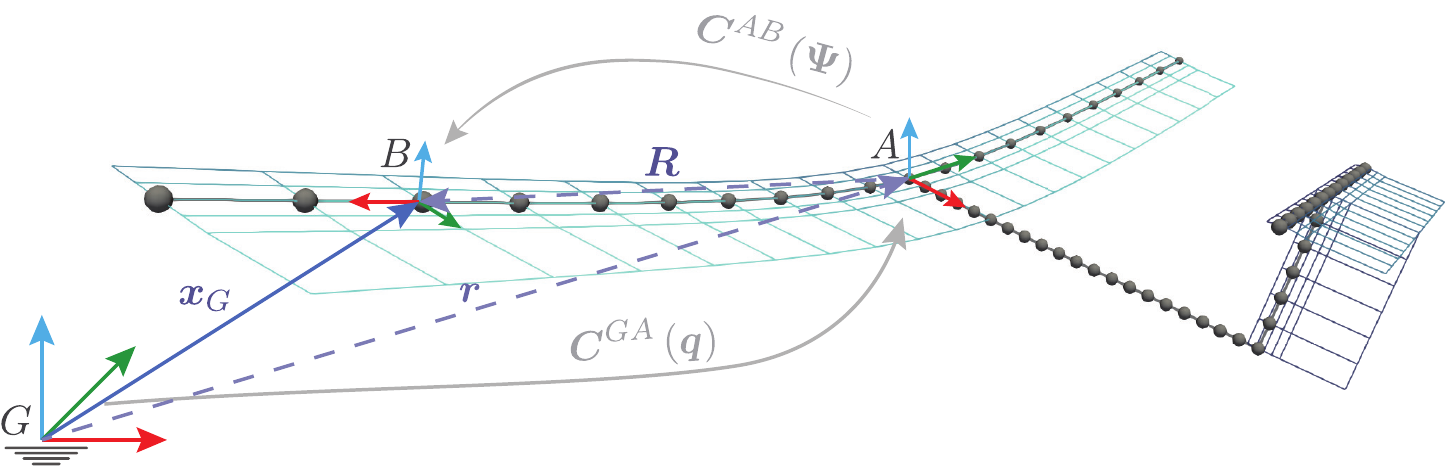}
\caption{Frames of reference used in the structural solver and parametrisation of the instantaneous geometry.}
\label{fig:frames_of_ref}
\end{figure*}

The aeroelastic model consists of a structural solver based on a non-linear geometrically exact beam theory (GEBT) and an aerodynamic model based on the unsteady vortex lattice method (UVLM), as originally proposed by \cite{MuruaEtAl2012}. The 1D structure is closely coupled to convergence at each FSI time step to the discrete-time UVLM using a coinciding spanwise panel discretisation. In particular, quadratic finite elements are used for the GEBT, with a discretisation based on nodal displacements and rotations, which are given in a body-attached, moving frame of reference $A$ (this frame moves and rotates with respect to an inertial frame, $G$) as shown in figure \ref{fig:frames_of_ref}.

The beam dynamics description is derived from the equations of a curved, geometrically nonlinear beam using the local element frame of reference $B$. Subsequently, the discrete system of equation (after applying a quadratic finite elements approximation) is expressed in terms of $\bm{\eta}$, the state variable containing the nodal displacements and rotations, and $[\bm{v}_A,\bm{\omega}_A]$, the state variable of velocities and rotations of frame $A$, as \citep{Hesse2016}
\begin{equation}
    \bm{\mathcal{M}}(\bm{\eta})
    \left\{ \begin{array}{c}
        \dot{\bm{v}}_A \\
        \dot{\bm{\omega}}_A
    \end{array}
    \right\} + \bm{\mathcal{Q}}_{\text{gyr}}(\bm{\eta}, \dot{\bm{\eta}}, \bm{v}_A,\bm{\omega}_A) + \bm{\mathcal{Q}}_\text{stiff}(\bm{\eta}) = \bm{\mathcal{Q}}_\text{ext}
\end{equation}
where $\bm{\mathcal{M}}$ is the mass matrix and $\bm{\mathcal{Q}}_{\text{gyr}}$, $\bm{\mathcal{Q}}_{\text{stiff}}$ and $\bm{\mathcal{Q}}_{\text{ext}}$ are the discrete gyroscopic, stiffness, and external generalised forces respectively. This equation is then solved iteratively using a fixed-point iteration scheme while time integration is carried out with an explicit Newmark-$\beta$ scheme formulated in incremental form. The applied forces are given in the material frame of reference ($B$) allowing for follower force input.

The aerodynamic solver is based on the unsteady vortex lattice method (UVLM), a three-dimensional nonlinear potential computational technique for lifting surfaces. To this end, vortex rings are placed on the wings of the aircraft and by applying the no-penetration condition, Biot-Savart's law and the Kutta-Joukowski condition we are able to calculate the bound circulation $\bf{\Gamma}$ vector used to calculate the force increment which is split into steady and unsteady (apparent mass) contributions. These two quantities can be integrated to yield the aerodynamic forces at the centre point of the segments 
\begin{equation}
    \partial \bm{F}_\text{steady} = \rho_\infty ||\bf{\Gamma}|| (\bm{v} \times \partial \bm{l}) \label{eq:steady}
\end{equation}
where $k \in [0, 4)$ is the segment counter and $\bm{v}$, $\partial \bm{l}$ the structure segment's velocity and length. The unsteady force at the centre point of the panel is calculated via, 
\begin{equation}
    \partial \bm{F}_\text{unsteady} = \rho_\infty \dot{||\bf{\Gamma}||}(t) (\hat{\bm{v}} \times \partial \bm{l}) \label{eq:unsteady}
\end{equation}
where $\hat{\bm{v}}$ is a panel-weighted velocity and $\dot{||\bf{\Gamma}||}(t)$ is the rate of change of the magnitude of bound vorticity with respect to an inertial observer.
To model the wake of the aircraft, a background-flow-convected wake model is used. In this way, we do not account for the mutual interactions between the free (shed) vortex rings, which significantly reduces the computational cost during each iteration \citep{CarrePalacios2019}. Instead, the wake is allowed to convect by the ambient flow and interact with the structure without exhibiting the near tip roll-up. Studies conducted by the authors have determined that an ambient flow-convected wake yields near identical results when compared to the free-force wake for the problems of interest, while the savings in computational cost are remarkable. The present mid-fidelity aeroelastic solver deals with the coupled behaviour of the structural/aerodynamic effects of flexible structures and it has been implemented within the framework \texttt{SHARPy}, written in python, with low level libraries in C++ and Fortran, and parallelisation using OpenMP. 

\subsection{3D velocity fields as aeroelastic inputs}
A final aspect of the present model is the methodology for combining the turbulence input data with the dynamic aeroelastic solver. Atmospheric turbulence fields either in a two-dimensional form (vertical planes) or three-dimensional sub-domains of a flow realisation where generated for a time period of \SI{120}{\second}. Subsequently, the data was stored and read by \texttt{SHARPy} at each time step. The LES input data was found to impose practical constraints in the reading and interpolation process as each snapshot amounts to more than 3 Gb of binary memory (for $1024 \times 512 \times 513 \sim \SI{270}\,$million grid points). This poses a bottleneck which may be approached in two ways. A direct solution is to read and store in RAM only the two necessary snapshots. This is suitable for distributed computers, where permanent storage latency is high, and nodes can be devoted to a small number of simulations each. However, shared memory computers usually have a lower permanent storage latency and the RAM available per processor node is lower. The solution is then to map the binary file stored in the relatively fast hard drive into an array-like structure, so data is directly read from the permanent storage. This solution was seen to drastically reduce the memory overhead with a 20\% increase of the interpolation time cost. The 3D velocity field reader implemented here supports both methods, and it implements a cache where data fields are read only once during each simulation. In the end, the decision of which method is used for the 3D field read is a combination of the architecture the software is being run in, the number of cases running simultaneously and the time between LES snapshots. Once the 3D field is mapped or copied into an array, a multivariable interpolation is conducted. First, the volumetric data is given as a structured regular grid. This makes trilinear interpolation a suitable solution, while the quicker closest neighbour interpolation is also available for finer 3D field discretisations. Results presented here are obtained using a trilinear interpolation in space. The time-dependency of the 3D velocity fields is considered in this analysis. When multiple snapshots of the field are given, a linear interpolation is performed between the two closest snapshots to the simulation time. A visualisation of the combined 3D velocity fields and aircraft mode is shown in figure \ref{fig:Combined3DFields_and_Aircraft} below.
\begin{figure}[ht]
	\includegraphics[width=\textwidth]{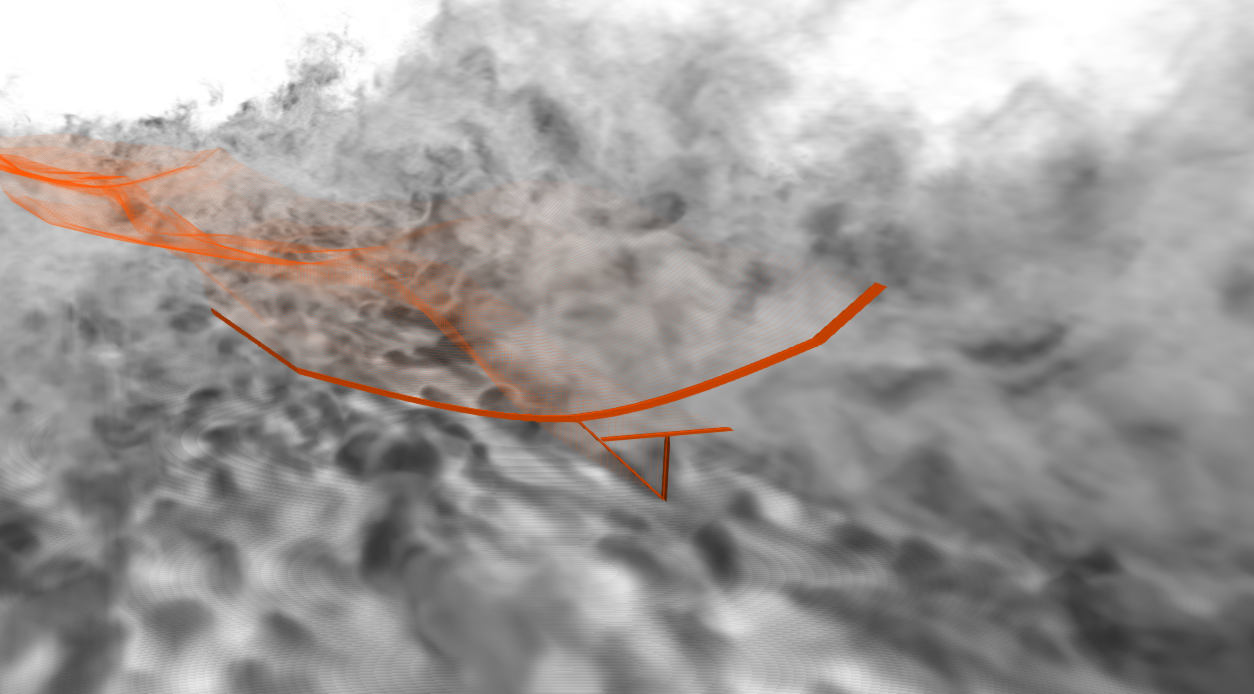}
	\caption{Visualisation of the 3D LES-generated velocity field (severe case) and its combination with the aircraft model are shown.}
	\label{fig:Combined3DFields_and_Aircraft}
\end{figure}
\section{Numerical results} \label{sec:AtmoAndAeroelast}
As discussed in Section \ref{sec:Intro}, in the design process of aircraft the 1D von K{\'a}rm{\'a}n spectrum is typically used to simulate continuous gusts. Having already presented the differences between the three turbulence-generation models (1D von K{\'a}rm{\'a}n, 2D Kaimal and 3D LES), in this section we shall attempt to quantify their effects on aircraft aeroelasticity from short simulated flights in open loop (that is, without feedback control on the vehicle dynamics) at the reference height $H_{\textnormal{ref}}=\,$\SI{10}{\metre}. 

\subsection{Test case definition}\label{sec:hale_t-tail}
Simulations will be carried out on the High-Altitude Long-Endurance (HALE) T-tail configuration of \cite{MuruaEtAl2012}, with minor modifications. This test case provides a simple, easily reproducible and representative HALE aircraft example, which has been previously used for aeroelastic simulations by \cite{MuruaEtAl2012} and \cite{Hesse2016}, among others. The vehicle features a high-aspect ratio wing with relatively low stiffness in torsion and out-of-plane bending and a conventional T-tail configuration, with elevator and rudder as control surfaces. All aircraft structures are flexible, although fuselage and tail are significantly stiffer than the wing. Finally, all aerodynamic surfaces of the aircraft are represented by a symmetric aerofoils. Two aircraft configurations are considered, which will be referred to as the \emph{stiff} and \emph{flexible} vehicles. They exhibit exactly the same geometry and inertial characteristics with the former stiffness properties scaled by a factor of 100. Details are presented in table \ref{table:HALETtailChar} for both configurations. The only active control surface in the simulations below is the elevator, and once the aircraft is trimmed for cruise flight, it is left fixed at that deflection.
\begin{figure*}[ht]
    \includegraphics[trim={0 6cm 10cm 5cm},clip,width=\textwidth]{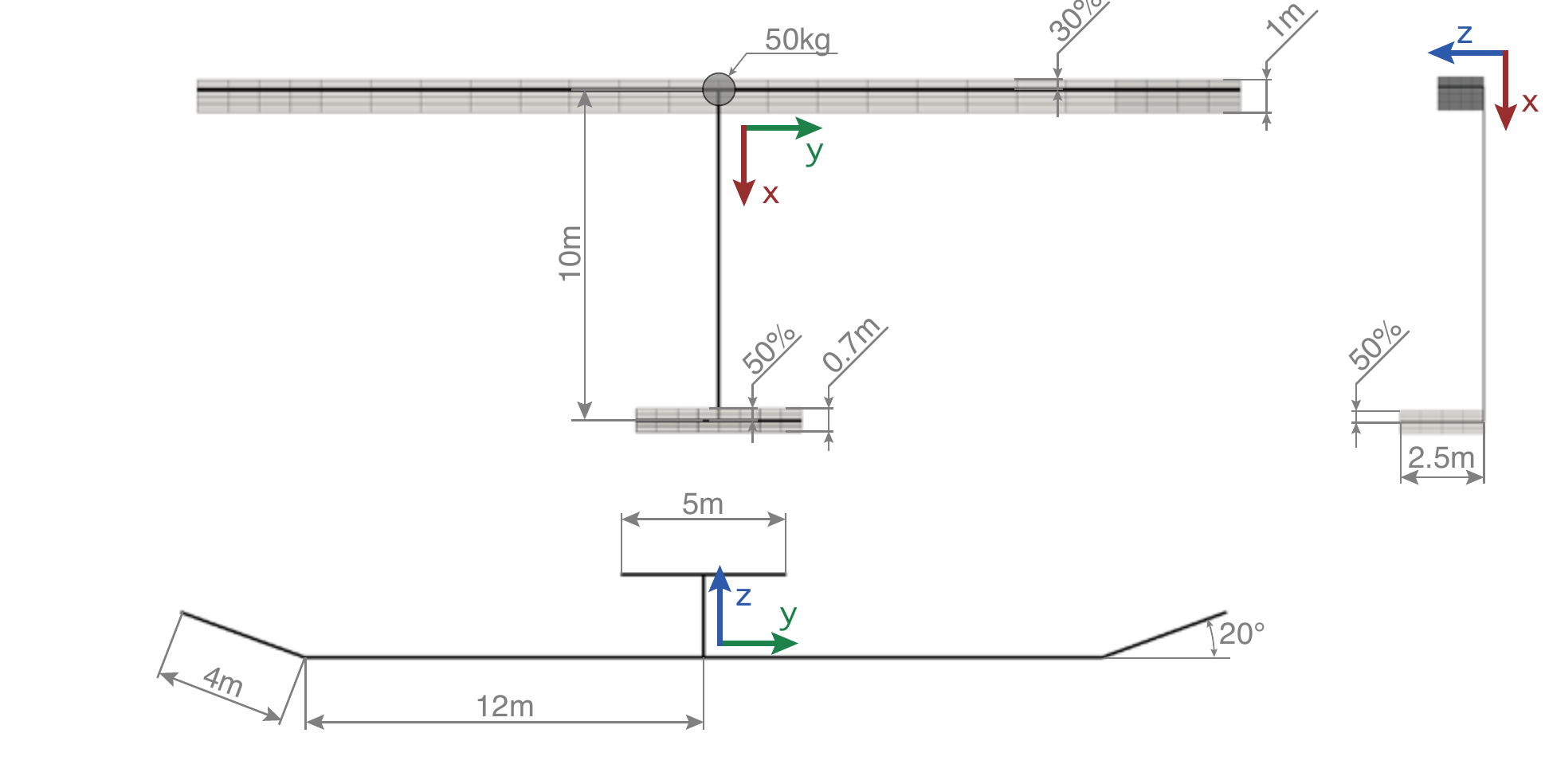}
    \caption{Schematic representation of the T-tail aircraft geometry.}
    \label{fig:HALE}
\end{figure*}
\begin{table*}[!h]
    \caption{\label{table:HALETtailChar} Stiffness and mass parameters for the aircraft model.}
    \centering
    \resizebox{\linewidth}{!}{\small
    \begin{tabular}{lcccccc}
        \hline
        Component & $EA$ $\left[\text{N}\cdot\text{m}\right]$ & $GJ$ $[\text{N}\cdot\text{m}^2]$  & $EI_y$ $GJ$ $[\text{N}\cdot\text{m}^2]$ & $EI_z$ $[\text{N}\cdot\text{m}^2]$ & $\bar{m}$ [kg$\cdot$m$^{-1}$]& $\bar{J}$ [kg$\cdot$m] \\ \hline
        & & & Flexible Aircraft& & &\\
        \hline
        Wing & 1.5e4&  1.5e4  &  6.0e4 &  4.2e6 & 0.75  & 0.1\\
        Fuselage & 1.5e5&  1.5e5  &  6.0e5 &  4.2e7  & 0.1  & 0.01\\
        Tail & 1.5e5&  1.5e5  &  6.0e5 &  4.2e7  & 0.1  & 0.01\\
        \hline
        & & & Stiff Aircraft & & &\\
        \hline
        Wing & 1.0e6 &  1.0e6  &  2.0e6 &  1.4e8 & 0.75  & 0.1\\
        Fuselage & 1.0e7 &  1.0e7  &  2.0e7 &  1.4e9  & 0.1  & 0.01\\
        Tail &1.0e7 &  1.0e7  &  2.0e7 &  1.4e9   & 0.1  & 0.01\\
        \hline
    \end{tabular}
    }
\end{table*}

\subsection{Flexible aircraft dynamics for each atmospheric turbulence model}\label{subsec:SensAircraftModel}
To examine the relative importance of input turbulence data in predicting both loads and flight trajectories for a flexible aircraft we undertake a number of simulations for all three turbulence intensity (TI) cases. Three-dimensional atmospheric turbulence fields are extracted from the three models such that they exhibit similar flow statistics, as discussed in section \ref{sec:CompTurbModels}, with simulations corresponding to \SI{120}{\second} of real time and $\Delta t=\,$\SI{0.025}{\second}. The produced time-series for the aircraft loads and trajectories for each TI and turbulence-generation model are subsequently used in our assessments. For the LES-generated background flows, multiple aeroelastic simulations were run for each turbulent intensity case in order to explore the spanwise flow variations of the simulated atmospheric turbulence. A summary of all aeroelastic simulations is shown in table \ref{table:SensitivitySimulations} below. Each case is defined by the average reference velocity and turbulence intensity as well as the initial spanwise position in the turbulent flow field, $Y$, of the aircraft mid-wing point. Values of $Y=\,$\SIlist{0;10;50;200;-200}{\metre} are selected based on the lateral turbulence integral length scale, which is calculated to be between \SIrange{150}{200}{\metre} for the three turbulence intensity cases.

\begin{table}[ht]
    \caption{\label{table:SensitivitySimulations} Summary of the various simulations along with some key parameters}
    \centering
    \begin{tabular}{lcccc}
        \hline
        Case & $U_{10}$ (m/s) & $\mathcal{I}_{10}$  (\%) & Y$(t=0)$ (m) & Turbulence model \\ \hline
        L11   & 5.05 & 8.87 & 0 & LES \\
        L12   & 5.05 & 8.87 & 10 & LES \\
        L13   & 5.05 & 8.87 & 50 & LES \\
        L14   & 5.05 & 8.87 & 200 & LES \\
        L15   & 5.05 & 8.87 & -200 & LES \\
        L21   & 5.41 & 10.74 & 0 & LES \\
        L22   & 5.41 & 10.74 & 10 & LES \\
        L23   & 5.41 & 10.74 & 50 & LES \\
        L24   & 5.41 & 10.74 & 200 & LES \\
        L25   & 5.41 & 10.74 & -200 & LES \\
        L31   & 4.74 & 12.73 & 0 & LES \\
        L32   & 4.74 & 12.73 & 10 & LES \\
        L33   & 4.74 & 12.73 & 50 & LES \\
        L34   & 4.74 & 12.73 & 200 & LES \\
        L35   & 4.74 & 12.73 & -200 & LES \\
        K1    & 5.05 & 8.87  & - & Kaimal \\
        K2    & 5.41 & 10.74 & - & Kaimal \\
        K3    & 4.74 & 12.73 & - & Kaimal \\
        VK1    & 5.05 & 8.87  & - & von K{\'a}rm{\'a}n \\
        VK2    & 5.41 & 10.74 & - & von K{\'a}rm{\'a}n  \\
        VK3    & 4.74 & 12.73 & - & von K{\'a}rm{\'a}n  \\
        \hline
    \end{tabular}
\end{table}

\begin{figure*}[ht]
    \includegraphics[width=\textwidth]{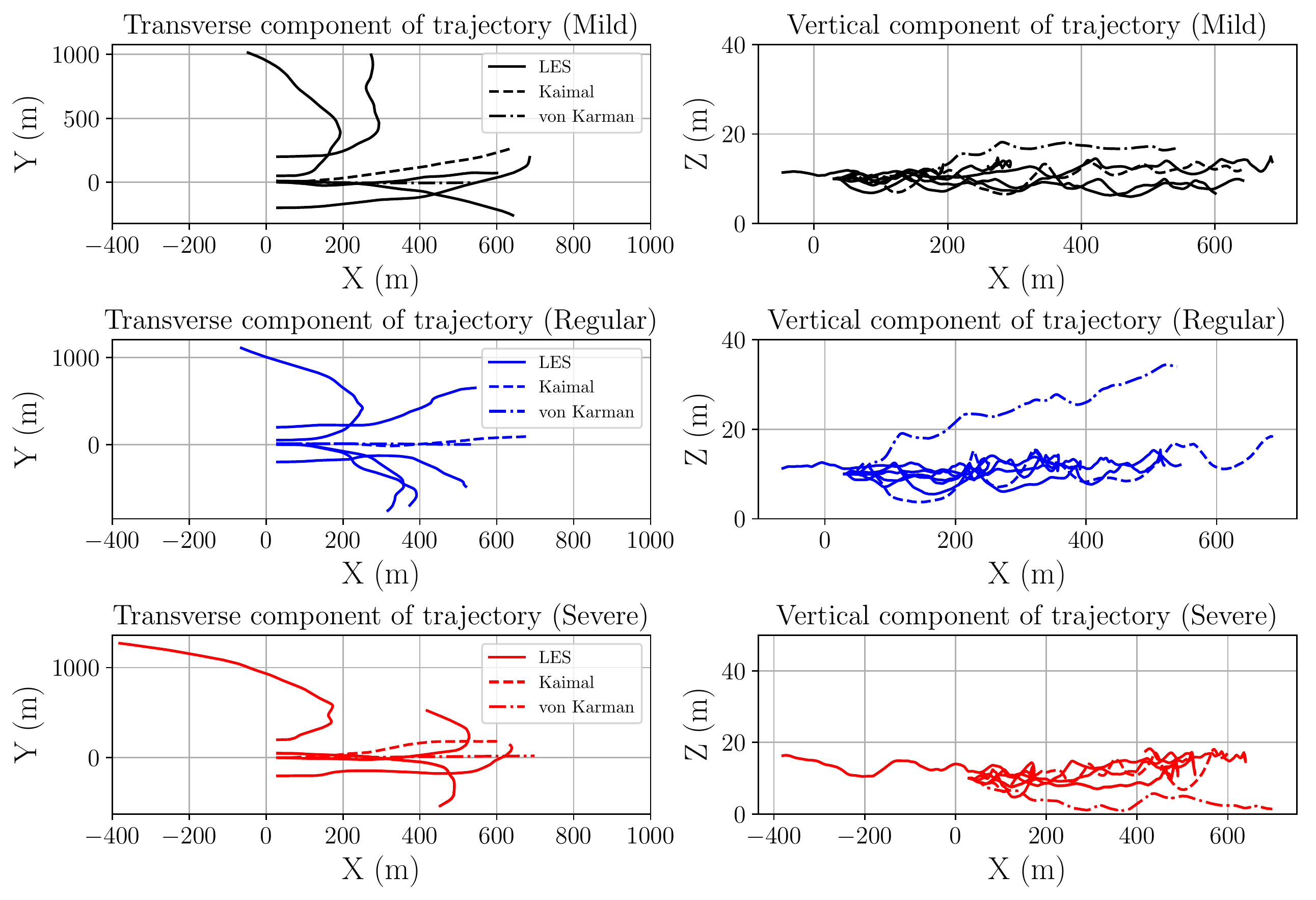}
    \caption{Flight trajectory of the flexible configuration and all three turbulence generation methods}
    \label{fig:FlightTrajectoriesTurbulenceModel}
\end{figure*}

\begin{figure*}[ht]
	\centering
    \includegraphics[trim={0cm 1cm 0cm 0cm},clip,width=0.8\textwidth]{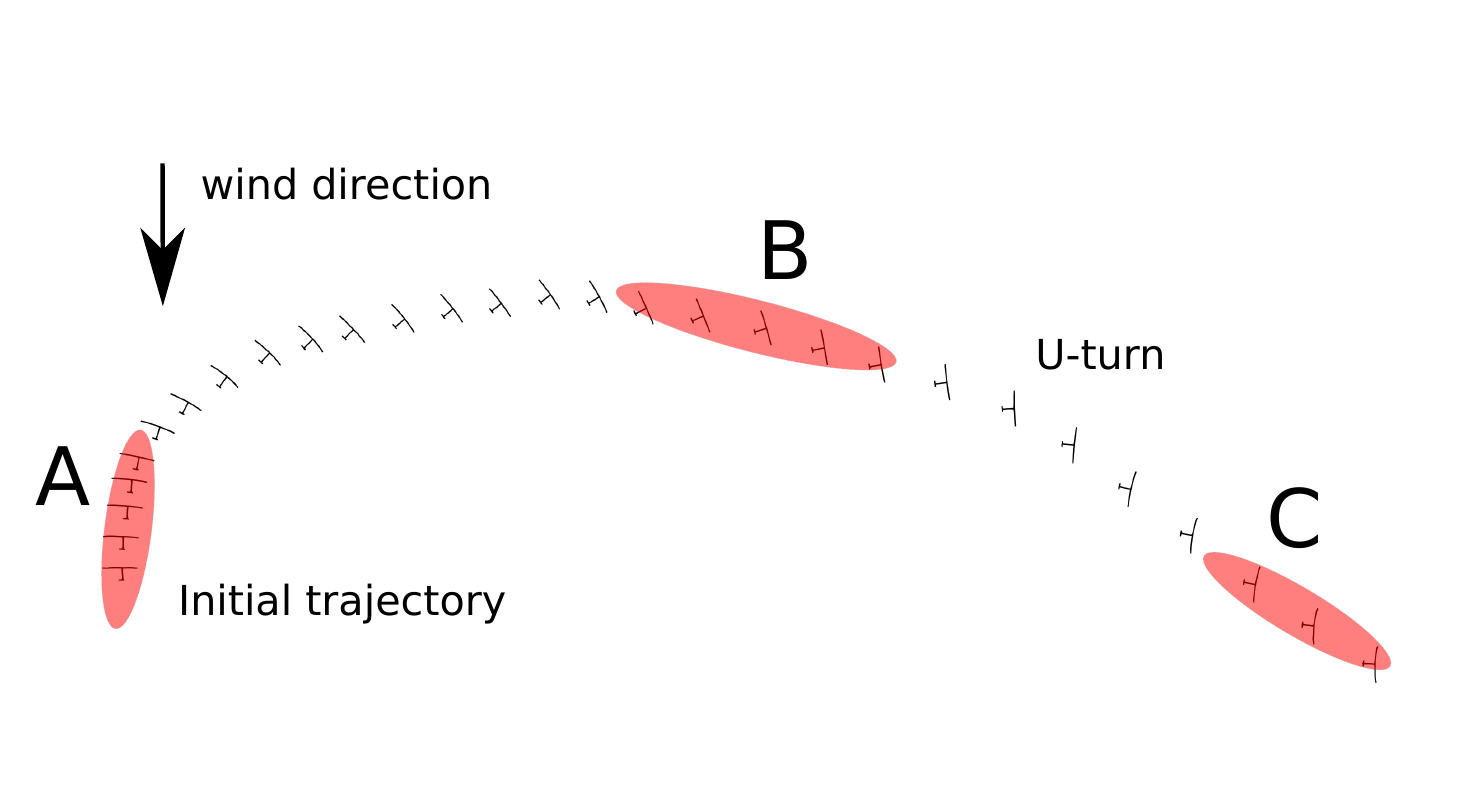}
    \includegraphics[width=0.3\textwidth]{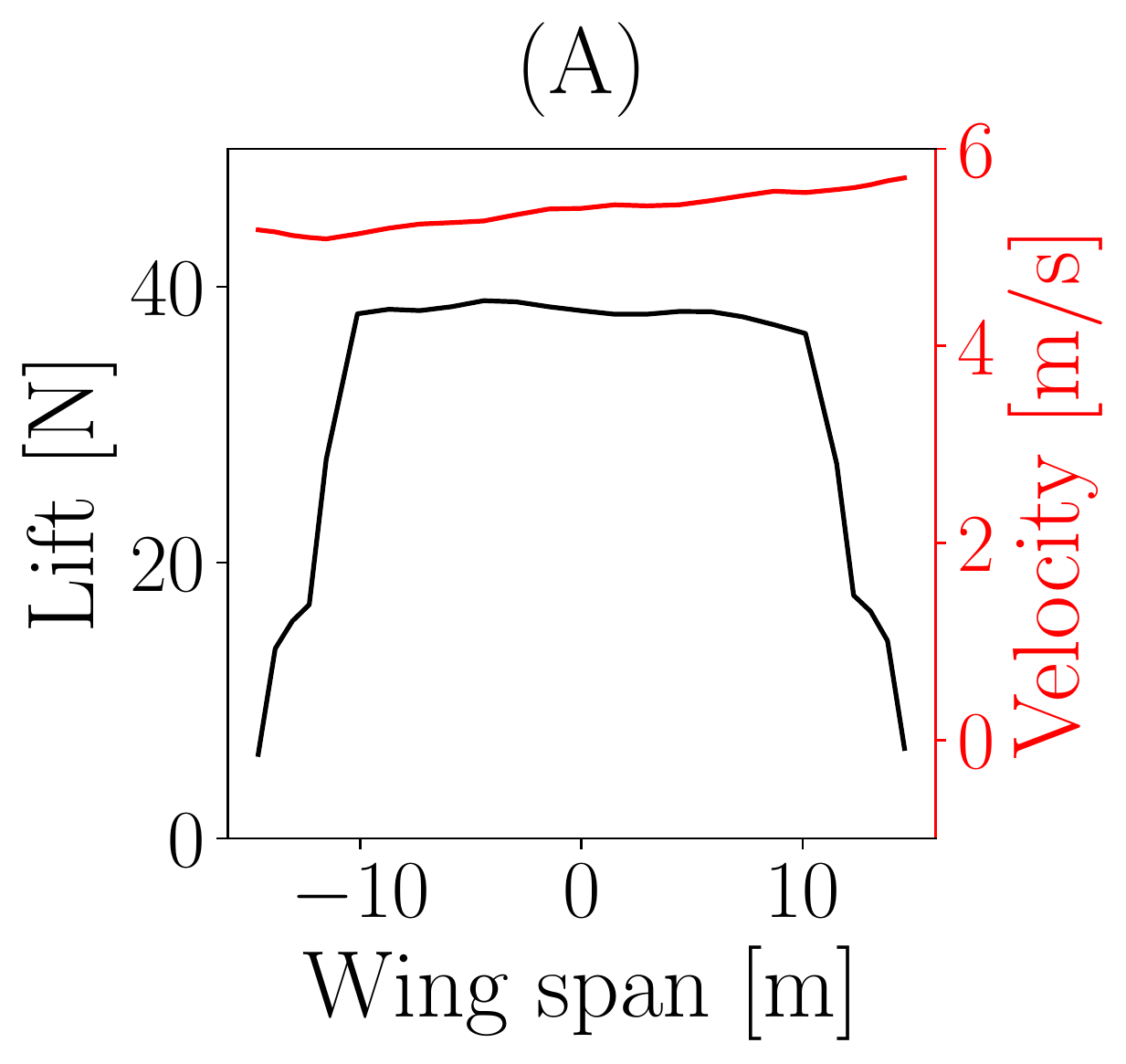}
    ~
    \includegraphics[width=0.3\textwidth]{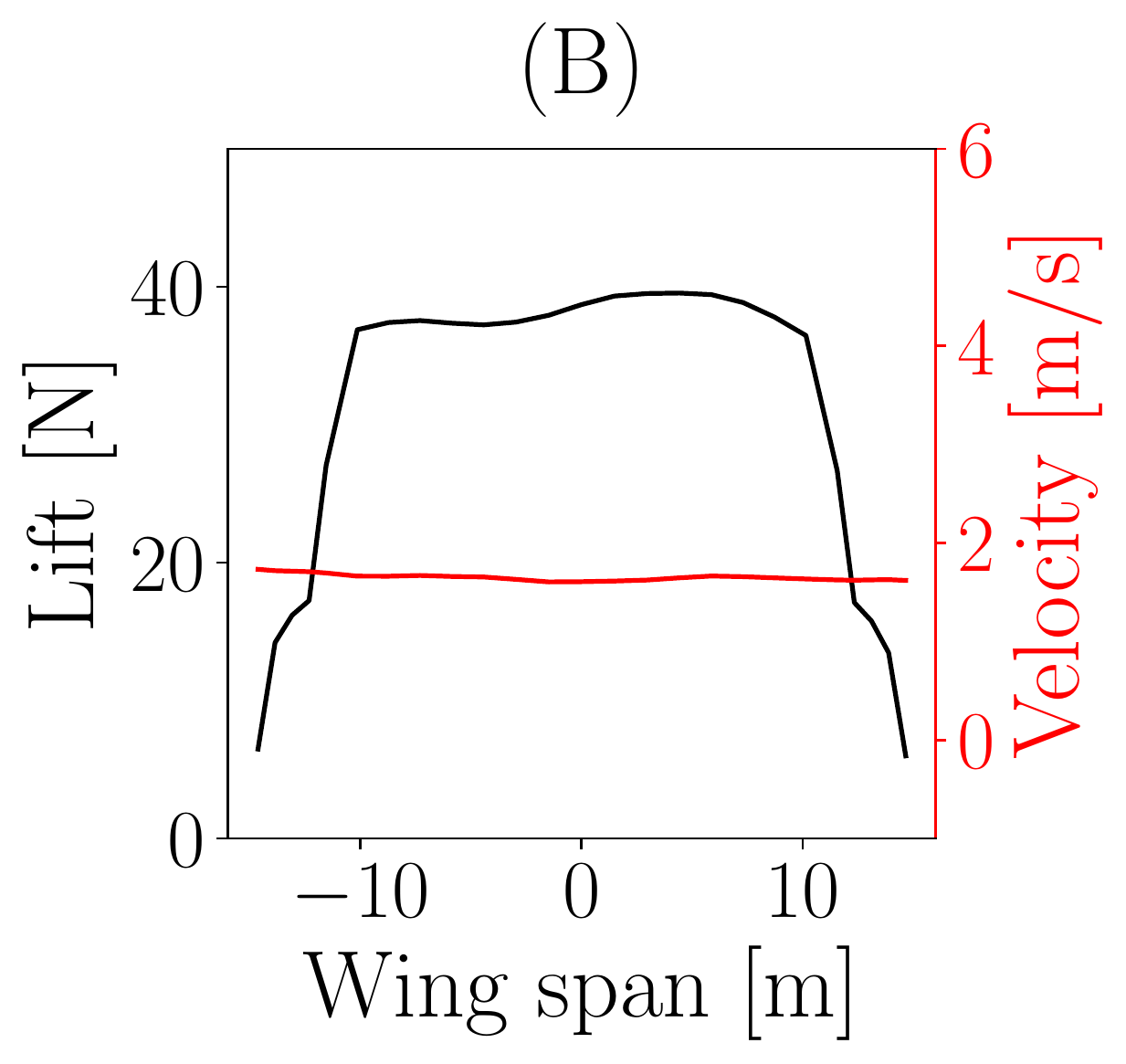}
    ~
    \includegraphics[width=0.3\textwidth]{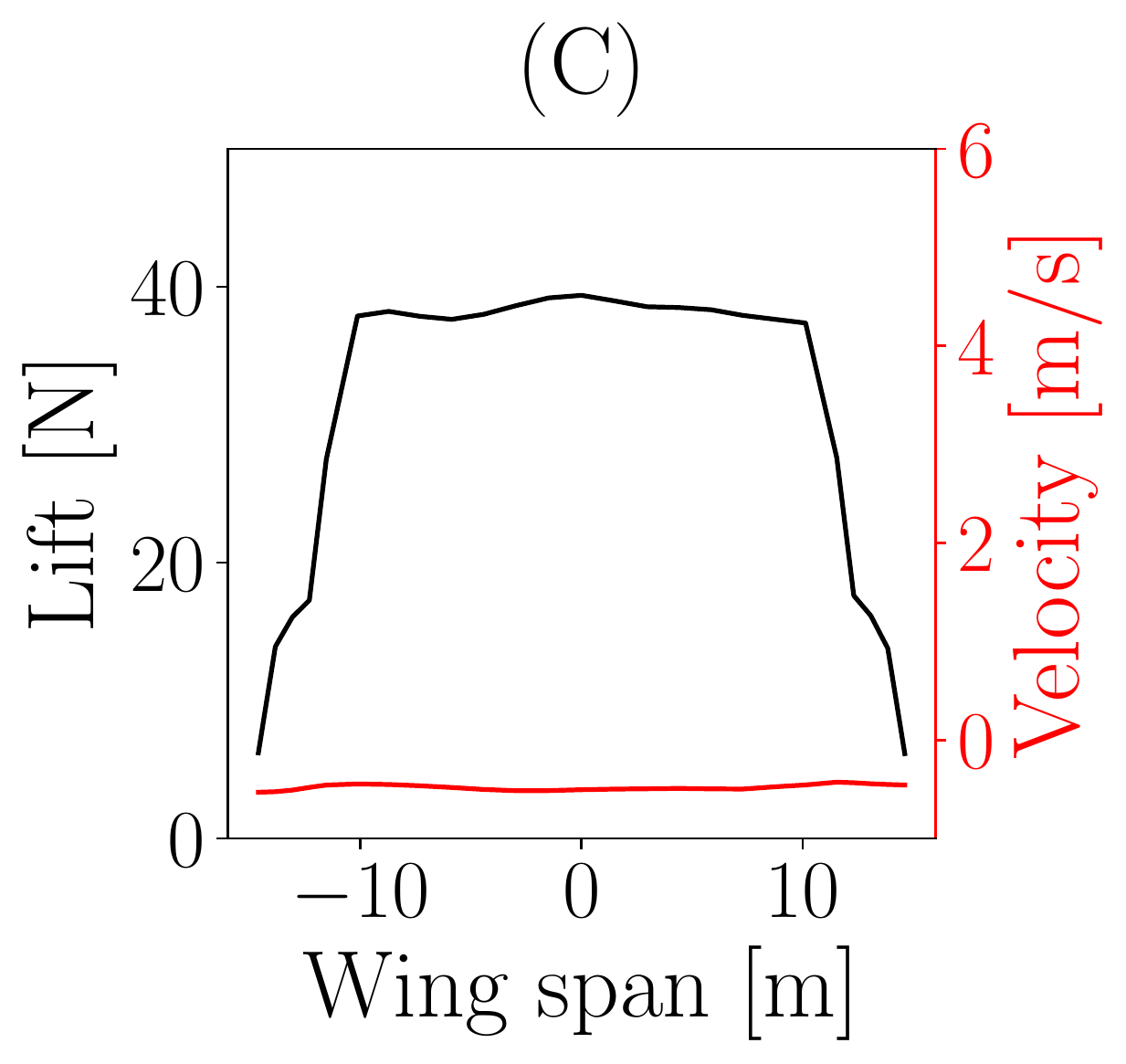}
    \caption{Flight trajectory and ensemble-averaged lift/velocity distribution of the flexible configuration for the regular LES case}
    \label{fig:FlightTrajectories_LiftDistributions}
\end{figure*}

\begin{figure*}[ht]
    \includegraphics[width=\textwidth]{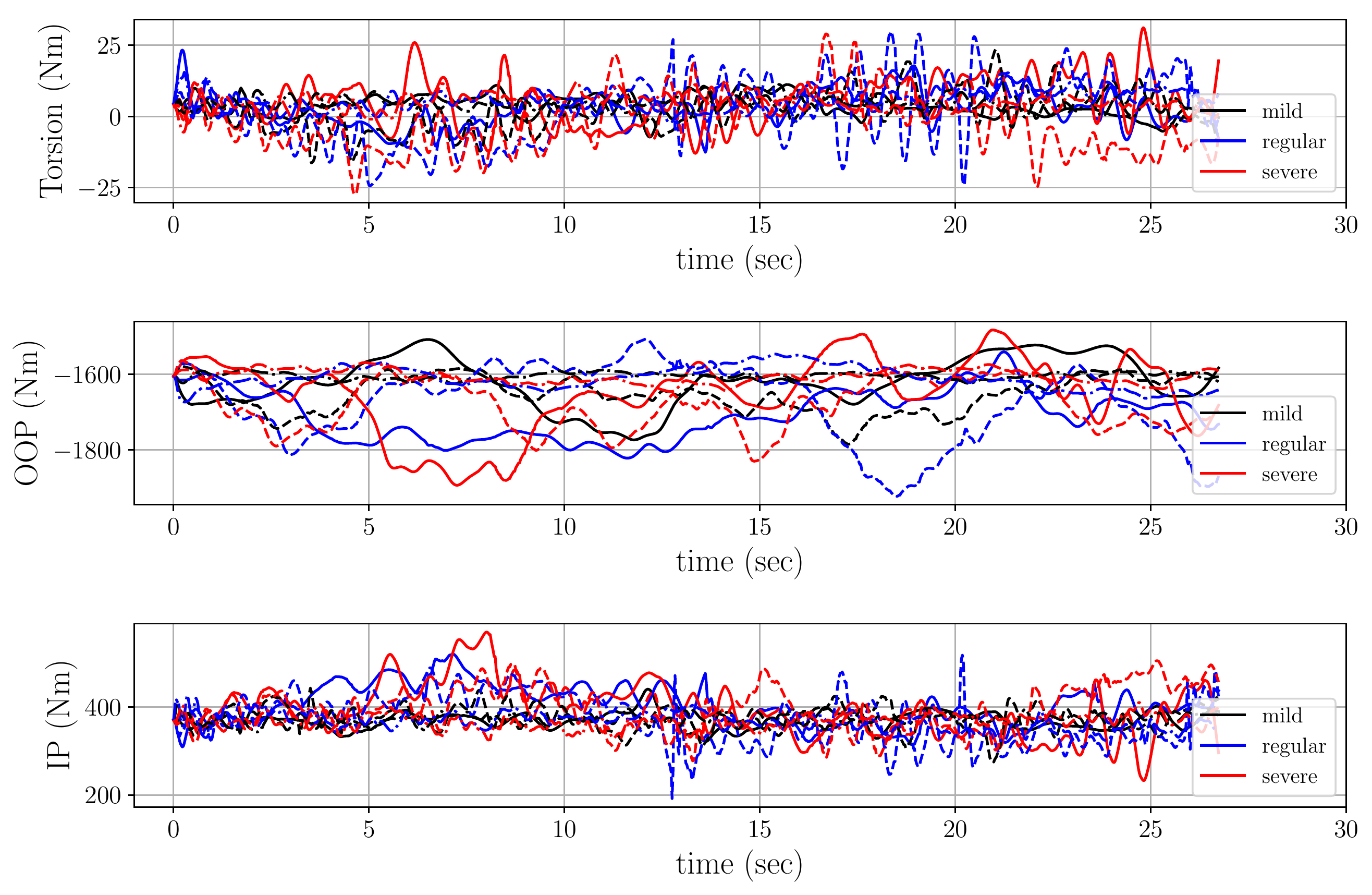}
    \caption{Root loads from the three turbulence generation models for the regular case. In the figure the torsion, out-of-plane (OOP) and in-plane (IP) moments are shown.}
    \label{fig:RootLoadsAllModels}
\end{figure*}

\begin{figure*}[ht]
    \includegraphics[width=\textwidth]{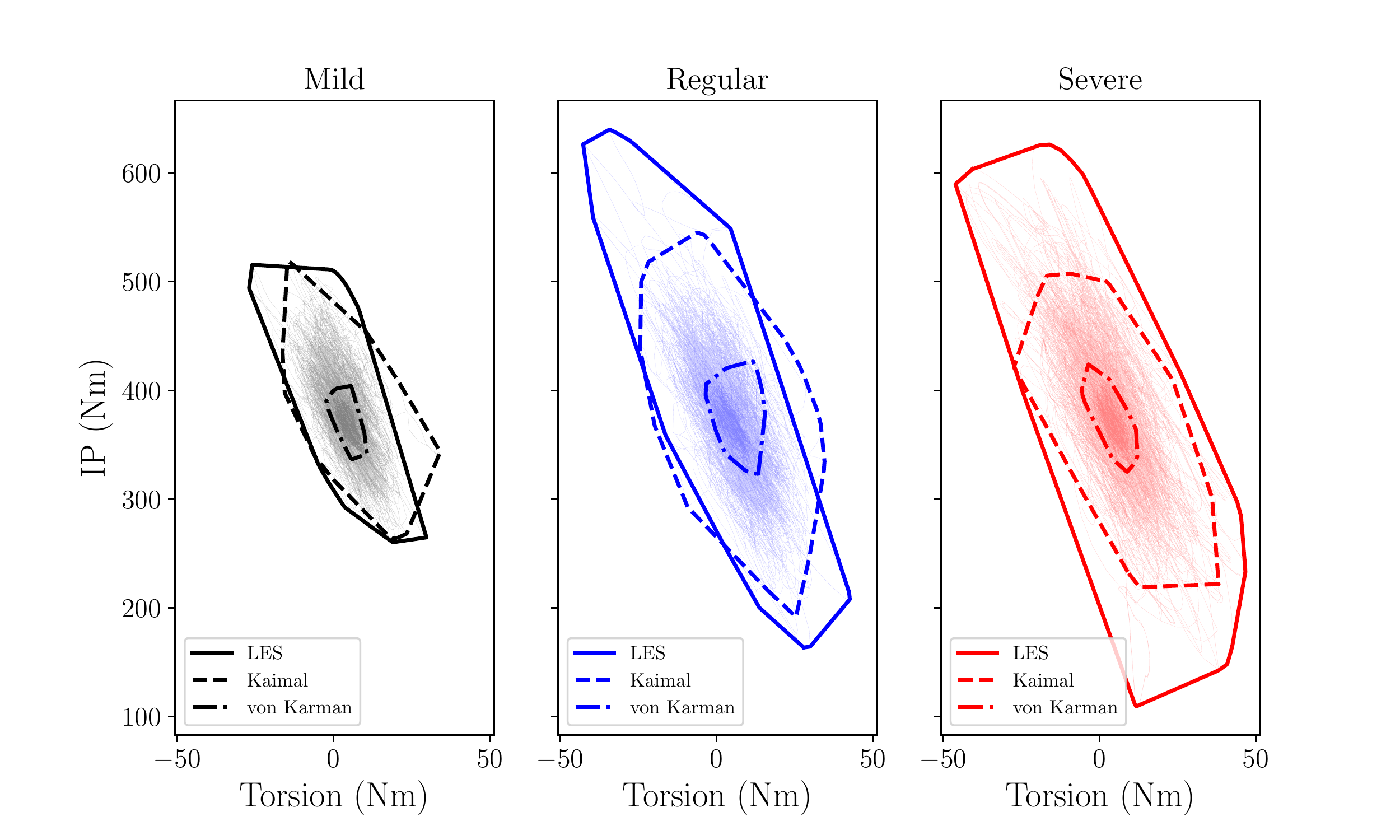}
    \caption{Wing root loads envelopes for all three models obtained from combining all available time histories.  In all figures the in-plane (IP) root moment is plotted against torsion.}
    \label{fig:RootLoadEnvelopesAllModels_IP}
\end{figure*}

Aircraft trajectories are tracked at the origin of the vehicle axes shown in figure \ref{fig:HALE} and the trajectories for all simulations in table \ref{table:SensitivitySimulations} are shown in figure \ref{fig:FlightTrajectoriesTurbulenceModel}. We may observe that the aircraft exhibits a similar vertical trajectory for all turbulence inputs by maintaining an altitude around \SI{10}{\metre} with ``dives'' and ``jumps'' of around \SIrange{1}{3}{\metre} occurring throughout. A significant deviation from this pattern is observed in the response to the von K{\'a}rm{\'a}n model for both the regular and severe cases. Perfect spanwise coherence in the background flow implies generation of net lift resultant that results in substantial inertia relief and an overall gain of altitude. 
The transverse component of the trajectory exhibits larger deviations between the three turbulence-generation models as well as within LES itself when the aircraft is initially placed at different spanwise locations with respect to the same background flow ($Y=\,$\SIlist{0;10;50;200;-200}{\metre}).  Interactions with the large scale flow features in the LES-generated turbulence results in significant variation of the aircraft trajectory, which can deviate both right and left while in some cases it may experience a full U-turn and continue flying in the opposite direction. The other two models (Kaimal and von K{\'a}rm{\'a}n), which are characterised by less or no variability in the transverse direction, result in nearly straight flight trajectories. 
To demonstrate the effect of spanwise velocity coherence experienced by the aircraft wing as well as its relationship with the aerodynamic behaviour (e.g. spanwise lift distribution) and aircraft trim shape, figure 
\ref{fig:FlightTrajectories_LiftDistributions} presents ensembled-averaged spanwsie velocity and lift distributions for the aircraft wing from an LES-generated turbulence case. This corresponds to the regular TI and the aircraft initiated at $Y=\,$\SI{50}{\metre} spanwise distance from the reference location (centre of the LES computational domain). For this case, a U-turn of the aircraft occurs during the simulation. 
Clearly, as the aircraft comes across large coherent structures it experiences an incident velocity with large variations across its span (see figure 
\ref{fig:FlightTrajectories_LiftDistributions}). In turn, the non-uniform velocity distribution gives rise to asymmetric loads which through the aircrafts flexibility lead to a gradual veer to the right, which leads to a vehicle U-turn. It should be noted that the aircraft altitude does not change significantly during the U-turn trajectory and that ensemble averaged lift distributions reduced further this variability. This is a result of a passive control mechanism inherent to the aircraft by its design which makes the aircraft navigate through turbulence without having full control over its trajectory. This of course can be addressed by a flight control system which can alter the aircraft trim and thus have an impact on loads as well. As far as the root loads are concerned, similar variability is observed for all three atmospheric turbulence inputs. In figure \ref{fig:RootLoadsAllModels} we observe that torsion fluctuates around a mean value of \SI{3.5}{\newton \metre} for all three models with larger amplitudes found for the regular and severe Kaimal cases. Out-of-plane (OOP) and in-plane (IP) wing root moments, on the other hand, attain a mean value of around \SI{-1700}{N \metre} and \SI{370}{N \metre}, respectively, with the fluctuations under the Kaimal and LES inputs being more pronounced in the time series. It is worth mentioning here that the aeroelastic simulations which make use of the Kaimal model exhibit relatively high-frequency fluctuations. These fluctuations which appear to dominate the loads after about \SI{12.5}{\second} coincide with the instance that the aircraft's trajectory deviates from a straight path (the one aligned with the mean flow). This effect can be attributed to the lack of streamwise coherence of the Kaimal model as well as the assumption of ``frozen turbulence'' which generates non-physical longitudinal structures once the 2D planes are projected to a three-dimensional wind field. Once the aircraft is misaligned from the mean flow, it is incident to a non-uniform velocity field with large velocity differences across its wing span, increasing thus the torsion experienced by the root. Similar fluctuations can also be observed for the ``severe'' case and the LES inputs. This however is not an artefact of the LES method but rather the effect of the near-surface large coherent flow structures. 
Nonetheless, in order to fully appreciate the magnitude of the loads it is worth plotting the load envelops for all three TI cases and root loads of interest, namely torsion, OOP and IP. 
\begin{figure*}[ht]
	\includegraphics[width=\textwidth]{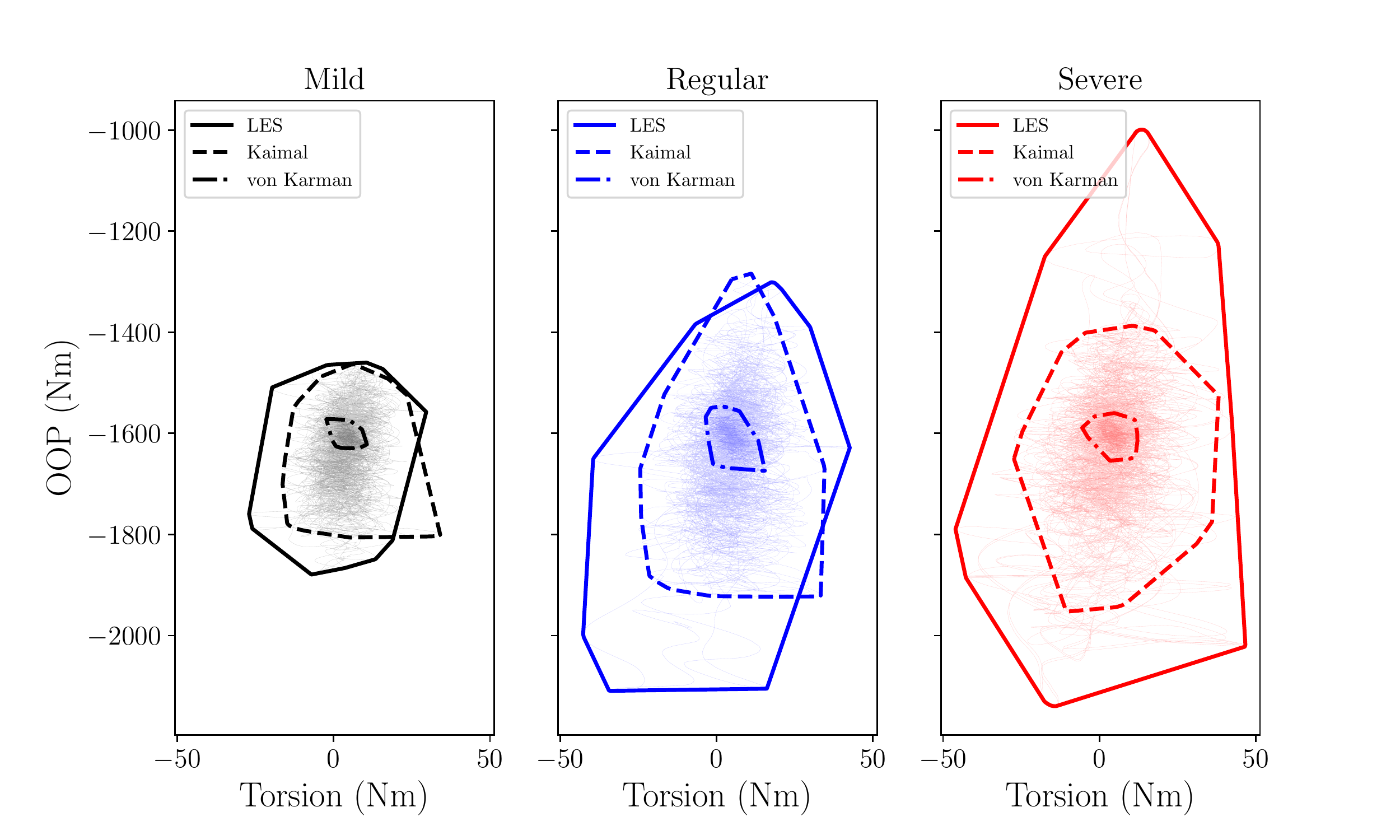}
	\caption{Root loads envelopes for all three models and cases.  In all figures the out-of-plane (OOP) root moment is plotted against torsion.}
	\label{fig:RootLoadEnvelopesAllModels_OOP}
\end{figure*}
The load envelops which are constructed by the convex hull obtained from the maximum instantaneous loads exerted on the aircraft, provide information regarding the loading regime expected for each TI case. To this end, from both figures \ref{fig:RootLoadEnvelopesAllModels_OOP} and \ref{fig:RootLoadEnvelopesAllModels_IP} we may observe the gradual expansion of the load envelopes as the input turbulence intensity increases. The shape of the load envelopes is also consistent in all three turbulence intensity cases. In the case of the OOP plotted against torsion, the load envelopes are stretched along the OOP direction signifying the significant increase of OOP with increasing levels of turbulence. Similarly, the shape of the IP versus torsion envelopes appears to be elongated with the larger in-plane moments occurring at negative root torsion. 
Nonetheless, in both cases (OOP and IP) the choice of the turbulence input model appears to be more significant. The LES data provide by far the largest load envelope with its boundaries twice as big as those of Kaimal. On the other hand, the Kaimal model yields smaller load envelopes with the difference between the two models being increased with the magnitude of turbulence intensity. Surprisingly, the smallest load envelopes are formed by the von K{\'a}rm{\'a}n model input. This is true for both the IP and OOP moments. We argue that this difference is due to the fact that in the von K{\'a}rm{\'a}n model, the spanwise coherence is by definition equal to unity, leading to a uniform spanwise load distribution. As a result, the dynamic loads give rise to a lifting of the overall structure through ``inertial relief'' (applied work by gust forces is transformed into rigid body kinetic energy). This is confirmed by figure \ref{fig:FlightTrajectoriesTurbulenceModel} in all three turbulence intensity cases by the vertical trajectory of the von K{\'a}rm{\'a}n model.
Another important difference that is captured by the load envelopes is the loads' frequency of occurrence. We may observe that the boundaries of all Kaimal and LES load envelopes are due to intermittent wind gusts which increase with increasing levels of turbulence. To determine the potential effects of intermittency on the aircraft loads, we have computed the probability distribution function (PDF) of the incident (background) velocity and load increments ($\Delta u=u(t+\Delta t)-u(t)$, $\Delta M=M(t+\Delta t)-M(t)$) and examined the characteristics of its ``tail''. Intermittency in both the incident velocity and the loads can be captured by heavy tail distributions that significantly deviate form the standard Gaussian distribution. 
\begin{figure*}[ht]
\includegraphics[width=\textwidth]{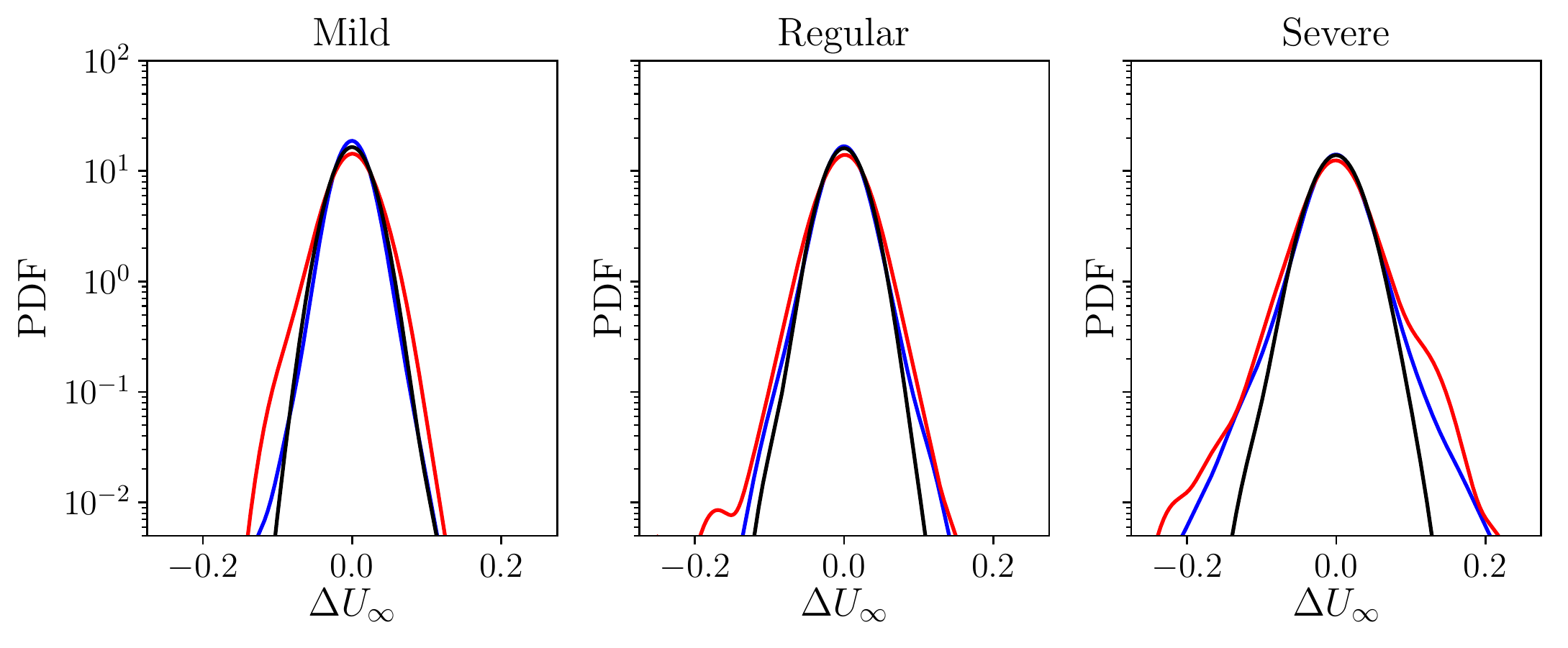}
\\
\includegraphics[width=\textwidth]{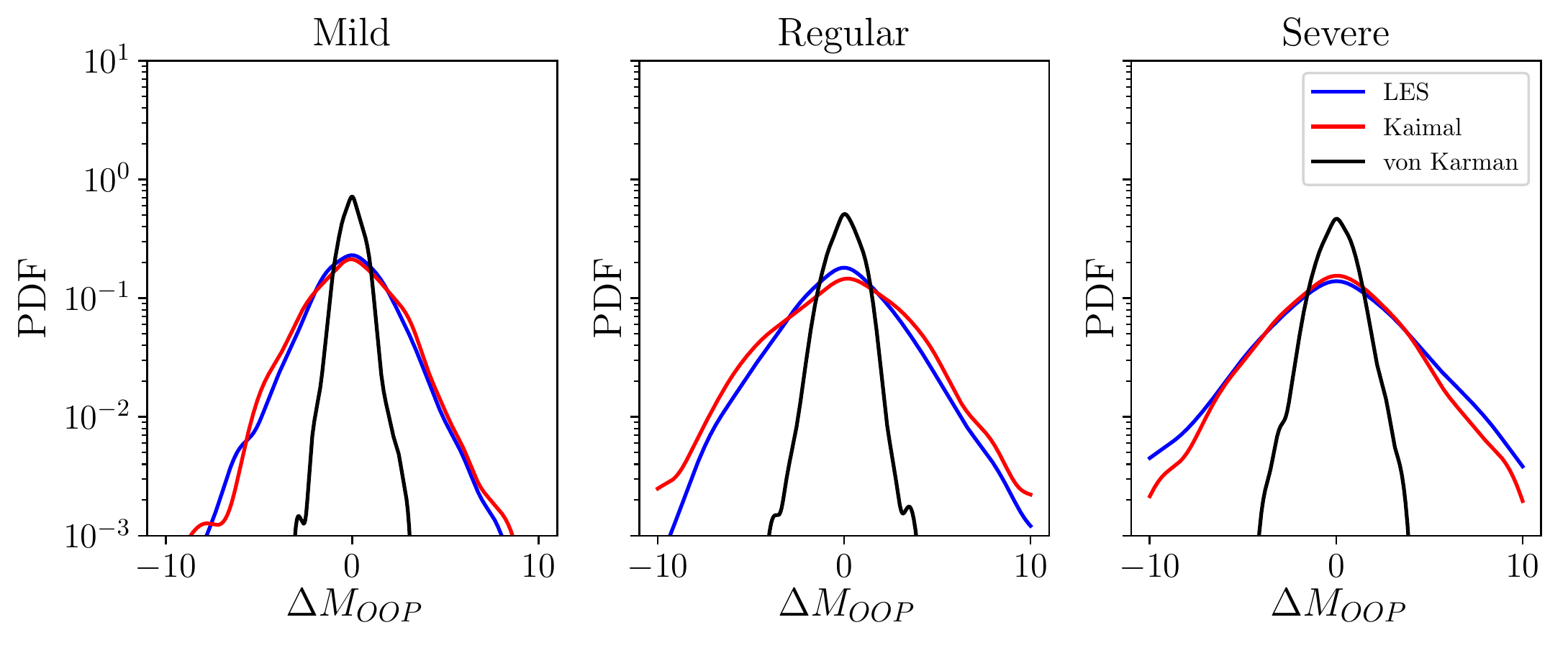}
\caption{PDF of the incident velocity magnitude$\Delta |U_\infty|$ and out-of-plane moment $M_{OOP}$ for all models and turbulence intensity cases. Normal distribution is included for comparison.}
\label{fig:Intermittency}
\end{figure*}
Figure \ref{fig:Intermittency} illustrates the existence of both intermittent velocity difference and loads, particularly for the ``severe'' case in which a clear heavy tail distributions, $|\Delta U_\infty|>\,$\SI{0.1}{\metre \per \second} and $|\Delta M_{OOP}|>\,$\SI{7.5}{\newton \metre} in both at positive and negative. It is worth noting that in the incident velocity difference PDFs the Kaimal model appears to be more intermittent. This is again due to the frozen turbulence assumption in combination with the deviation of the aircraft from a straight path. The wind field intermittency also translates into the observed loads. More specifically, the PDFs for $|\Delta M_{OOP}|$ show that the aircraft experiences large variation of out-of-plane loads sporadically with probability density of less that \num{0.01}, particularly in the ``severe'' case. In the other two ``milder'' cases the two models (Kaimal and LES) appear to be in a better agreement. Von K{\'a}rm{\'a}n turbulence is characterised by significantly less variability and thus, its PDF appears much narrower.  
\begin{figure*}[!ht]
\centering
\includegraphics[width=0.95\textwidth]{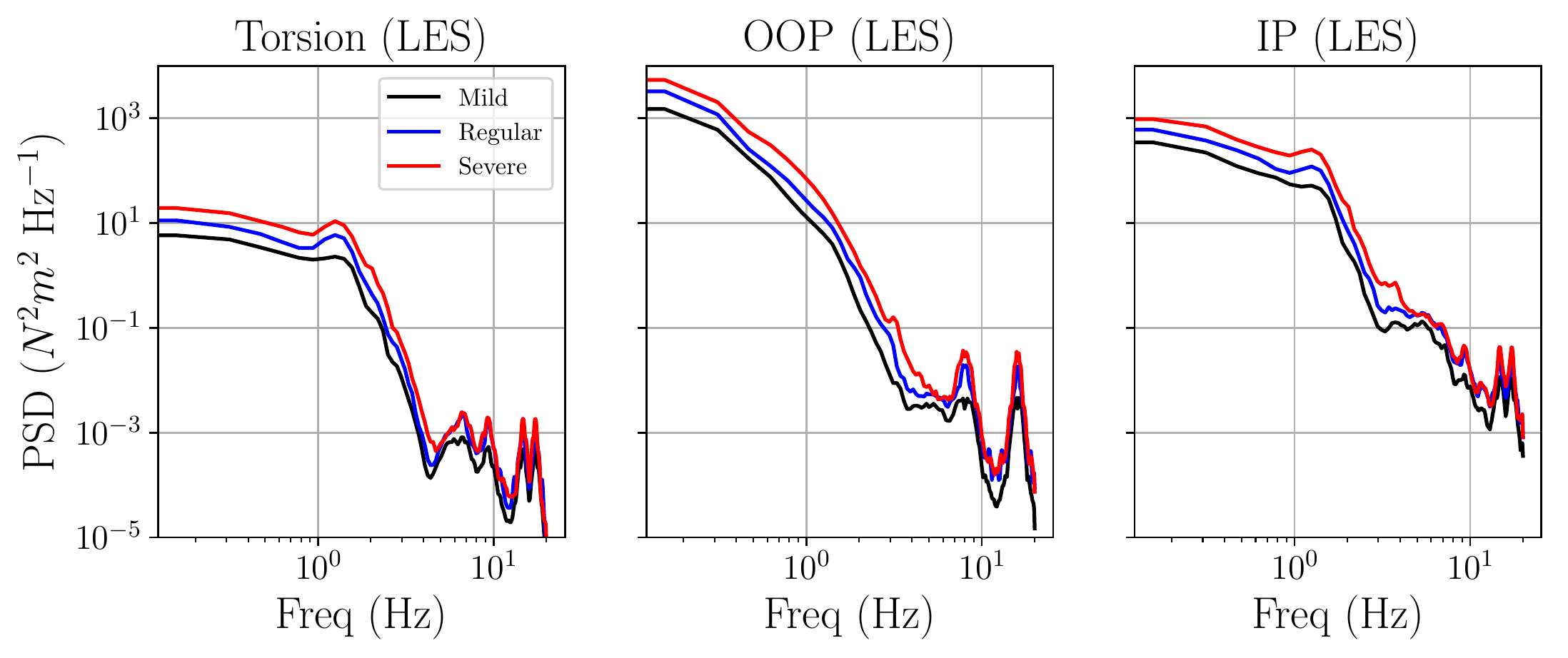}
\\
\includegraphics[width=0.95\textwidth]{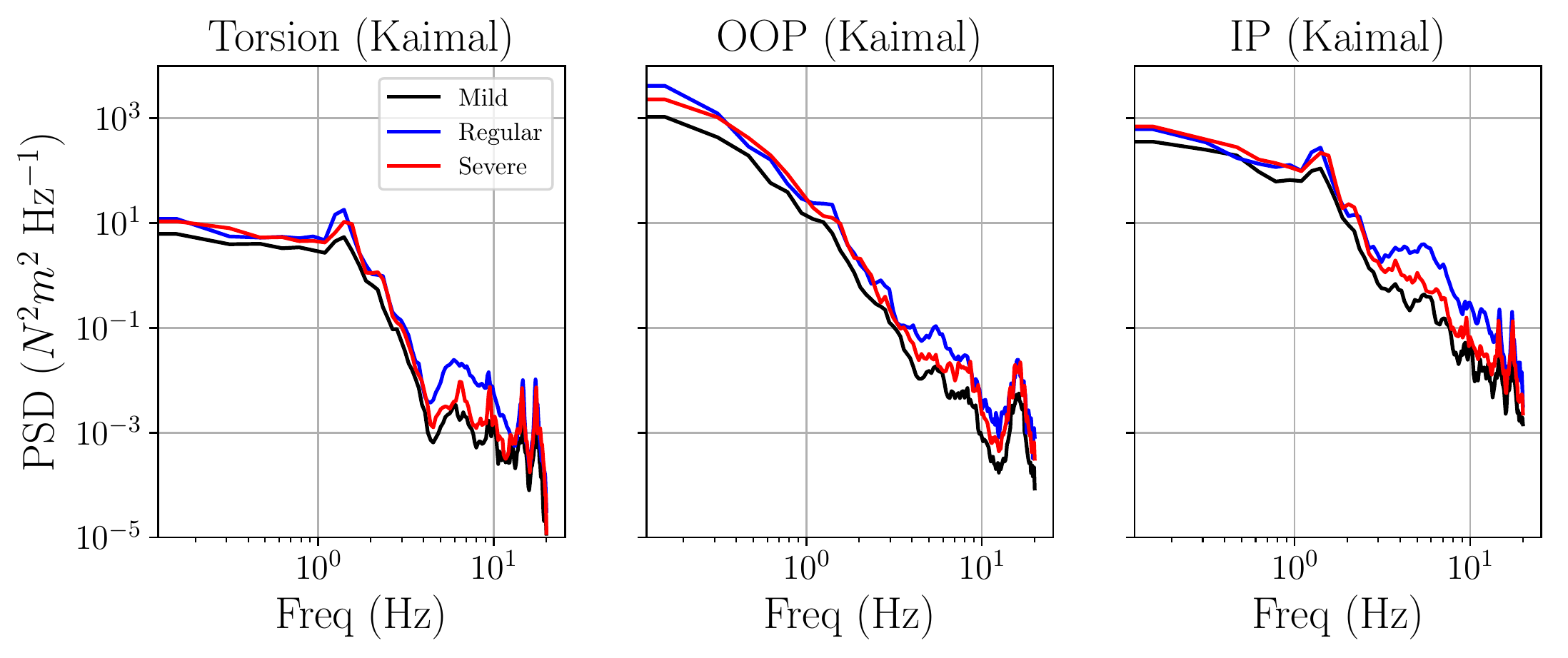}
\\
\includegraphics[width=0.95\textwidth]{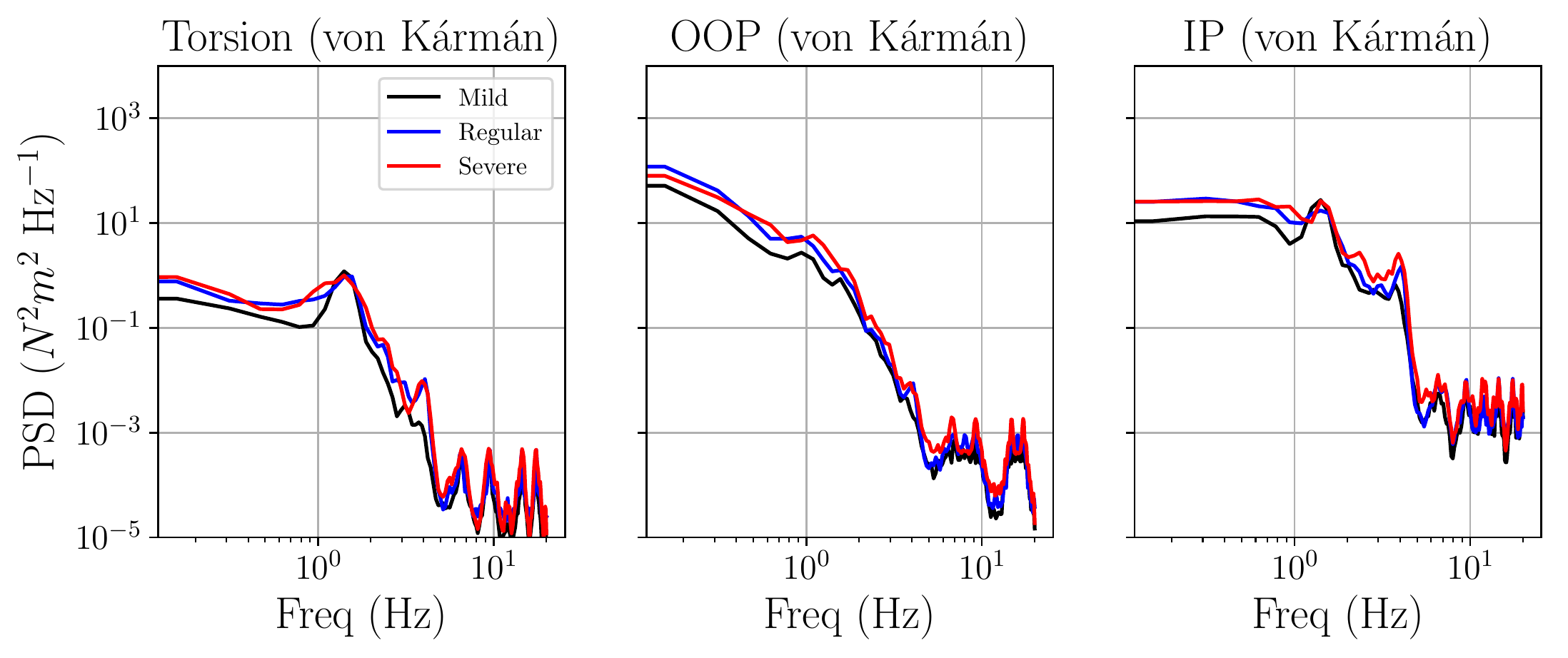}
\caption{PSDs of the root loads (Torsion, OOP and IP) for all three turbulence input cases (LES, Kaimal and von K{\'a}rm{\'a}n).}
\label{fig:LoadsSpectraAllmodels}
\end{figure*}
Comparing the OOP moment distribution with the other two models as well as with that of the incident velocity for the von K{\'a}rm{\'a}n model we may not be able to infer the origin of such a large difference. However, following our earlier argument we may again justify such a large difference in the PDF occurs by the stronger inertia relief in this case. Finally, it is worth presenting the power spectral density (PSD) functions of the root loads from all the three models in figure \ref{fig:LoadsSpectraAllmodels}. Again, comparing the PSDs for the three models a similar behaviour can be obtained. For instance, the PSD of torsion is characterised by a plateau region over of low-frequency region, followed by a steep slope and low-energy content $\sim$\num{0.001} in the high frequencies. In addition, all three cases exhibit a peak around $f=\,$\SI{1.3}{\hertz} which corresponds to the mode II (natural torsion frequency). The peak around mode II is more pronounced for the severe turbulence case while modes I ($f=\,$\SI{0.4}{\hertz}) and V ($f=\,$\SI{3.92}{\hertz}) are less recognisable in the PSD plots. Nonetheless, the effect of increasing the level of turbulence intensity on each PSD can be seen in all three cases by ``pushing'' the distributions up. This effect is clearer over the lower frequencies (\SIrange{0.1}{1}{\hertz}) and in particular for the simulations using the LES-generated turbulence input. A similar trend is followed by the IP moment. On the other hand, the OOP follows a smoother slope and lacks a ``plateau'' region, confirming that its response varies across the range of scales. Nonetheless, it is again worth noting that again the von K{\'a}rm{\'a}n simulations yield PSD distributions with an amplitude of about an order of magnitude smaller than their respective counterparts generated by LES and Kaimal turbulence inputs.  
\subsection{Aeroelastic effects assessment}
\begin{figure*}[ht]
    \includegraphics[width=\textwidth]{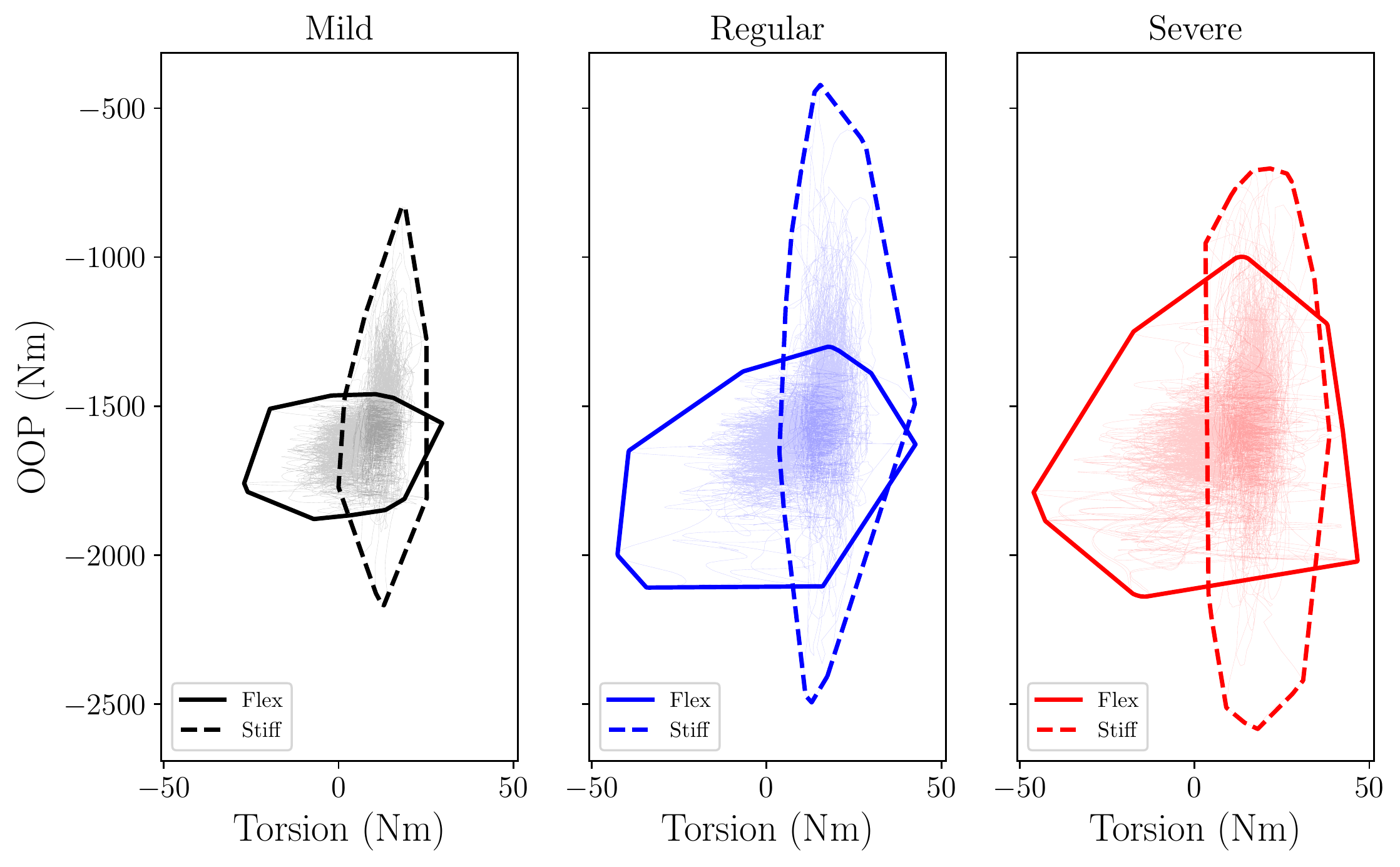}
    \caption{Root loads envelopes for the ``Stiff'' and ``Flexible'' cases for all levels of TI. In all figures the out-of-plane (OOP) root moment is plotted against torsion.}
    \label{fig:RootLoadEnvelopes_StiffVsFlex_OOP}
\end{figure*}

Having demonstrated the differences between three turbulence input models (von K{\'a}rm{\'a}n, Kaimal and LES), in this section we present a cross-comparison between the \emph{Flexible} and the \emph{Stiff} aircraft configurations. For this analysis we only use the LES turbulence inputs. The two aircraft configurations are expected to exhibit a different response to dynamic gust loading. The flight dynamics of the former (\emph{Flexible}) aircraft configuration are characterised by its coupling with the low-frequency structural modes \cite{Afonso2017}. On the other hand, the \emph{Stiff} aircraft configuration is expected to exhibit a substantial degree of decoupling between the two. To address the question of how much does stiffness affect the flight dynamics and structural loads of an aircraft subject to atmospheric turbulence, we present and compare the load envelopes and PSD distributions of the two configurations. To this end, the two aircraft configurations have been simulated for the same number of cases by maintaining all other parameters the same. We begin with the presentation of the OOP vs torsion load envelope in figure \ref{fig:RootLoadEnvelopes_StiffVsFlex_OOP}.  
Aircraft flexibility is found to fundamentally change the dynamic gust response of the aircraft. The stiff aircraft is found to be prone to larger OOP root moment loads and suffer less from large variations in torsion. More specifically, OOP moment attains values ranging from \SI{-500}{\newton \metre} to \SI{-2500}{\newton \metre} while experiencing only positive torsion values throughout. 
\begin{figure*}[ht]
    \includegraphics[width=\textwidth]{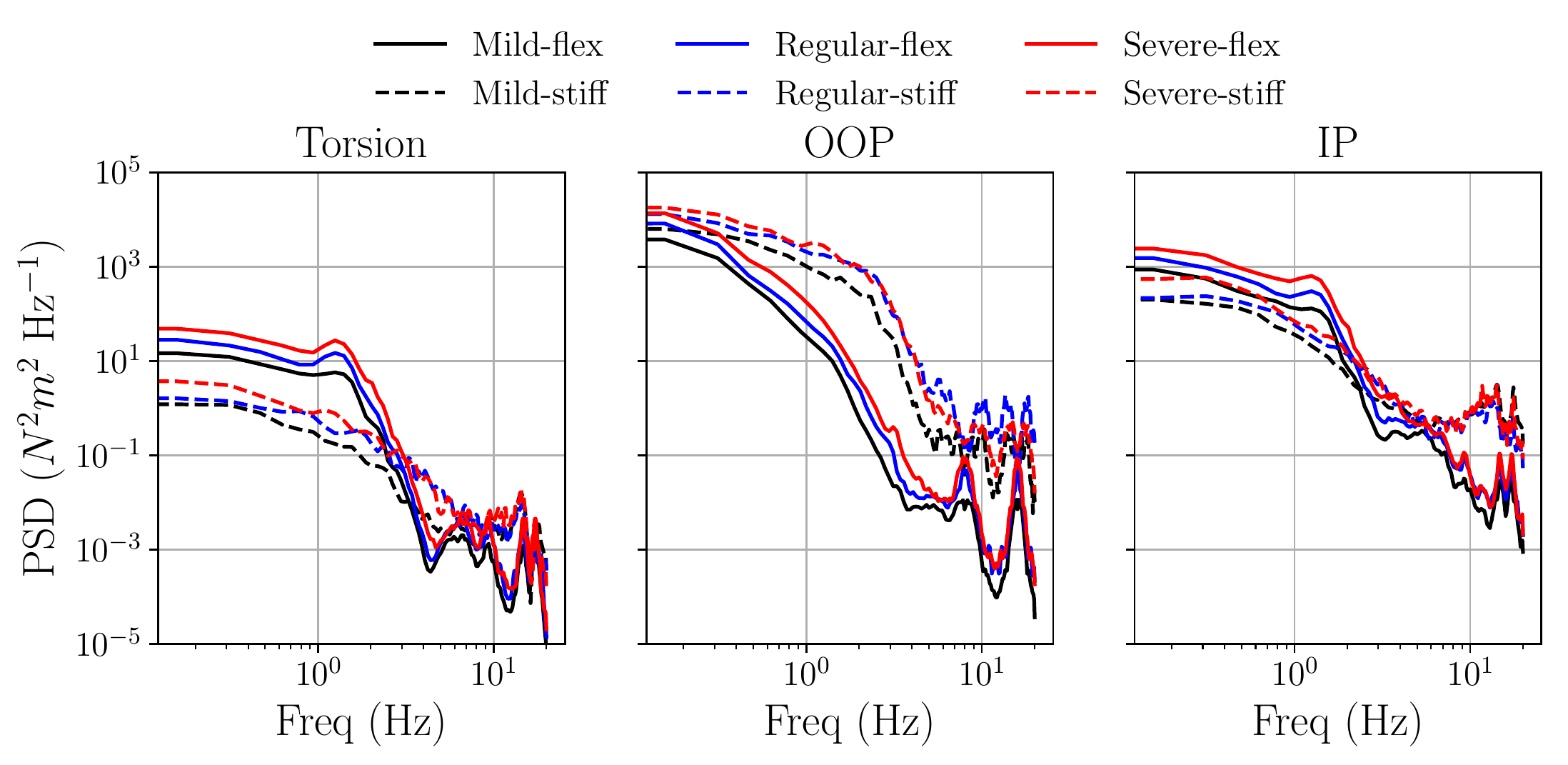}
    \caption{Comparison of PSD distributions for the root loads (Torsion, OOP and IP) between the stiff and flexible aircraft.}
    \label{fig:LoadsSpectra_StiffVsFlex}
\end{figure*}
 This shift in the loads response is also reflected in the respective PSDs of the loads. Figure \ref{fig:RootLoadEnvelopes_StiffVsFlex_OOP} shows a comparison between the two cases for all computed root loads. As in the load envelopes, torsion fluctuations of the stiff case are an order-of-magnitude) smaller than those of the flexible one with the differences being more pronounced in the low-frequency region. This is a well-known steady aeroelastic effect that cascades in -- lift twists the wing thus reducing the bending moments and increases torsion. Conversely, the OOP PSD shows a significant increase in the energy over the mid-range frequencies (inertial subrange). Finally smaller difference can be observed for the IP moment with the stiff case exhibiting smaller fluctuations. This transfer of energy from torsion and IP moment fluctuations to OOP signifies the fundamental differences in the turbulence/loads transfer function between the flexible and the stiff aircraft. The former, transforms the OOP loads into torsion which in turn changes its aerodynamic behaviour and therefore the resultant root loads. This can be seen as a passive load alleviation mechanism which targets a reduction of the OOP moment. The stiff aircraft on the other hand, by experiencing a certain degree of decoupling between structural and flight dynamics, remains more responsive to incident turbulence (the $-5/3$ scaling is maintained in the inertial sub-range) implying it may experience larger fatigue loads during its operation. 
\section{Conclusions}\label{sec:Conclusions}
In the present study we have conducted aircraft aeroelasticity simulations using turbulence input fields which are assumed to be representative of their operating environment. The aeroelastic coupling and dynamic loading on a very flexible aircraft was found to be highly modulated by the coherent structures and flow features of low-altitude atmospheric turbulence. This may be of critical importance for some very light air vehicles currently under consideration. 

The choice of the turbulence generation method was found to have a significant impact on both loads and flight dynamics of the aircraft. LES and Kaimal models provide similar results with LES being able to predict larger load envelopes. Kaimal was found to be problematic when it is used within our moving frame of reference as the application of the frozen turbulence hypothesis used to transform temporal 2D planes into spatial full three-dimensional wind fields does not generate three-dimensional coherence in the wind field. As a result, small deviations of the aircraft from the dominant wind direction generated sudden ``load jumps'' which affected the load envelopes in our simulations. LES, on the other hand, simulates a more realistic turbulent environment with large scales extending in the streamwise direction and generating flow coherence. The von K{\'a}rm{\'a}n model was the worst performing method of those considered, particularly when used in a moving three-dimensional frame of reference. The uniform planar velocity fields from this approach result in spanwise loading which substnunderpredicts the magnitude of the loads and overpredicts the energy transferred into rigid body motion. Therefore its use for loads and aeroelastics at low altitudes is not advised.

The comparison between the flexible and the stiff cases has demonstrated the extent of aeroelastic effects in the previous simulations. Analysis of the two aircraft flight dynamics shows that stiffer aircraft are prone to instabilities from high-frequency perturbations (gusts) leading to large OOP moment loads. More flexible vehicles, on the other hand, have natural load alleviation mechanisms through structural (elastic) and rigid body motion coupling. Results have shown smaller OOP loads as the energy contained in gust loading is distributed between rigid body motion, torsion, OOP and IP moments. Overall, as the flight dynamics of high-aspect ratio flexible aircraft were found to be fundamentally different from that of the stiff one, more emphasis should be given to non-linear aeroelastic phenomena which might occur under turbulent conditions and are currently not well-captured by existing computational models. 

\section*{Acknowledgements}
The authors would like to acknowledge funding from EPSRC (grant EP/R007470/1) and Airbus Defence and Space.
\section*{References}
\bibliographystyle{model1-num-names}
\bibliography{Bibfile.bib}

\end{document}